# Pathogenesis, Symptomatology, and Transmission of SARS-CoV-2 through Analysis of Viral Genomics and Structure

**This in progress manuscript is not intended for the general public.** This is a review paper that is authored by scientists for an audience of scientists to discuss research that is in progress. If you are interested in guidelines on testing, therapies, or other issues related to your health, you should not use this document. Instead, you should collect information from your local health department, the CDC's guidance, or your own government.

# Authors

- **Halie M. Rando** 0000-0001-7688-1770 rando2 tamefoxtime  Department of Systems Pharmacology and Translational Therapeutics, University of Pennsylvania, Philadelphia, Pennsylvania, United States of America; Department of Biochemistry and Molecular Genetics, University of Colorado School of Medicine, Aurora, Colorado, United States of America; Center for Health AI, University of Colorado School of Medicine, Aurora, Colorado, United States of America · Funded by the Gordon and Betty Moore Foundation (GBMF 4552); the National Human Genome Research Institute (R01 HG010067)

- **Adam L. MacLean** 0000-0003-0689-7907 adamlmaclean adamlmaclean  Department of Quantitative and Computational Biology, University of Southern California, Los Angeles, California, United States of America

- **Alexandra J. Lee** 0000-0002-0208-3730 ajlee21  Department of Systems Pharmacology and Translational Therapeutics, University of Pennsylvania, Philadelphia, Pennsylvania, United States of America · Funded by the Gordon and Betty Moore Foundation (GBMF 4552)

- **Ronan Lordan** 0000-0001-9668-3368 RLordan el_ronan  Institute for Translational Medicine and Therapeutics, Perelman School of Medicine, University of Pennsylvania, Philadelphia, PA 19104-5158, USA

- **Sandipan Ray** 0000-0002-9960-5768 rays1987  Department of Biotechnology, Indian Institute of Technology Hyderabad, Kandi, Sangareddy 502285, Telangana, India

- **Vikas Bansal** 0000-0002-0944-7226 bansalvi VikasBansal1989  Biomedical Data Science and Machine Learning Group, German Center for Neurodegenerative Diseases, Tübingen 72076, Germany

- **Ashwin N. Skelly** 0000-0002-1565-3376 anskelly  Perelman School of Medicine, University of Pennsylvania, Philadelphia, Pennsylvania, United States of America; Institute for Immunology, University of Pennsylvania Perelman School of Medicine, Philadelphia, United States of America · Funded by NIH Medical Scientist Training Program T32 GM07170

- **Elizabeth Sell** 0000-0002-9658-1107 esell17  Perelman School of Medicine, University of Pennsylvania, Philadelphia, Pennsylvania, United States of America

- **John J. Dziak** 0000-0003-0762-5495 dziakj1  Edna Bennett Pierce Prevention Research Center, The Pennsylvania State University, University Park, PA, United States of America

- **Lamonica Shinholster** 0000-0001-6285-005X LSH2126  Mercer University, Macon, GA, United States of America · Funded by the Center for Global Genomics and Health Equity at the University of Pennsylvania

- **Lucy D'Agostino McGowan** 0000-0001-7297-9359 LucyMcGowan LucyStats  Department of Mathematics and Statistics, Wake Forest University, Winston-Salem, North Carolina, United States of America

- **Marouen Ben Guebila** 0000-0001-5934-966X marouenbg marouenbg  Department of Biostatistics, Harvard School of Public Health, Boston, Massachusetts, United States of America

- **Nils Wellhausen** 0000-0001-8955-7582 nilswellhausen  Department of Systems Pharmacology and Translational Therapeutics, University of Pennsylvania, Philadelphia, Pennsylvania, United States of America

- **Sergey Knyazev** 0000-0003-0385-1831 Sergey-Knyazev SeKnyaz  Georgia State University, Atlanta, GA, United States of America

- **Simina M. Boca** 0000-0002-1400-3398 SiminaB  Innovation Center for Biomedical Informatics, Georgetown University Medical Center, Washington, District of Columbia, United States of America

- **Stephen Capone** 0000-0001-7231-1535 scapone01  St. George's University School of Medicine, St. George's, Grenada

- **Yanjun Qi** 0000-0002-5796-7453 qiyanjun  Department of Computer Science, University of Virginia, Charlottesville, VA, United States of America

- **YoSon Park** 0000-0002-0465-4744 ypar **yoson** Department of Systems Pharmacology and Translational Therapeutics, University of Pennsylvania, Philadelphia, Pennsylvania, United States of America · Funded by NHGRI R01 HG10067

- **Yuchen Sun** kevinsunofficial Department of Computer Science, University of Virginia, Charlottesville, VA, United States of America

- **David Mai** 0000-0002-9238-0164 davemai daveomai Department of Bioengineering, University of Pennsylvania, Philadelphia, PA, USA

- **Joel D Boerckel** 0000-0003-3126-3025 jboerckel jboerckel Department of Orthopaedic Surgery, Perelman School of Medicine, University of Pennsylvania, Philadelphia, PA, United States of America; Department of Bioengineering, University of Pennsylvania, Philadelphia, PA, United States of America

- **Christian Brueffer** 0000-0002-3826-0989 cbrueffer cbrueffer Department of Clinical Sciences, Lund University, Lund, Sweden

- **James Brian Byrd** 0000-0002-0509-3520 byrdjb thebyrdlab University of Michigan School of Medicine, Ann Arbor, Michigan, United States of America · Funded by NIH K23HL128909; FastGrants

- **Jeremy P. Kamil** 0000-0001-8422-7656 Department of Microbiology and Immunology, Louisiana State University Health Sciences Center Shreveport, Shreveport, Louisiana, USA

- **Jinhui Wang** 0000-0002-5796-8130 jinhui2 Perelman School of Medicine, University of Pennsylvania, Philadelphia, Pennsylvania, United States of America

- **Ryan Velazquez** 0000-0002-3655-3403 rdvelazquez Azimuth1, McLean, Virginia, United States of America

- **Gregory L Szeto** 0000-0001-7604-1333 gregszetoAI greg_szeto Allen Institute for Immunology, Seattle, WA, United States of America

- **John P. Barton** 0000-0003-1467-421X johnbarton _jpbarton Department of Physics and Astronomy, University of California-Riverside, Riverside, California, United States of America

- **Rishi Raj Goel** 0000-0003-1715-5191 rishirajgoel rishirajgoel Institute for Immunology, University of Pennsylvania, Philadelphia, PA, United States of America

- **Serghei Mangul** 0000-0003-4770-3443 smangul1 serghei_mangul Department of Clinical Pharmacy, School of Pharmacy, University of Southern California, Los Angeles, CA, United States of America

- **Tiago Lubiana** 0000-0003-2473-2313 lubianat lubianat  Department of Clinical and Toxicological Analyses, School of Pharmaceutical Sciences, University of São Paulo, São Paulo, Brazil

- **COVID-19 Review Consortium**

- **Anthony Gitter** 0000-0002-5324-9833 agitter anthonygitter  Department of Biostatistics and Medical Informatics, University of Wisconsin-Madison, Madison, Wisconsin, United States of America; Morgridge Institute for Research, Madison, Wisconsin, United States of America · Funded by John W. and Jeanne M. Rowe Center for Research in Virology

- **Casey S. Greene** 0000-0001-8713-9213 cgreene GreeneScientist  Department of Systems Pharmacology and Translational Therapeutics, University of Pennsylvania, Philadelphia, Pennsylvania, United States of America; Childhood Cancer Data Lab, Alex's Lemonade Stand Foundation, Philadelphia, Pennsylvania, United States of America; Department of Biochemistry and Molecular Genetics, University of Colorado School of Medicine, Aurora, Colorado, United States of America; Center for Health AI, University of Colorado School of Medicine, Aurora, Colorado, United States of America · Funded by the Gordon and Betty Moore Foundation (GBMF 4552); the National Human Genome Research Institute (R01 HG010067)

**COVID-19 Review Consortium:** Vikas Bansal, John P. Barton, Simina M. Boca, Joel D Boerckel, Christian Brueffer, James Brian Byrd, Stephen Capone, Shikta Das, Anna Ada Dattoli, John J. Dziak, Jeffrey M. Field, Soumita Ghosh, Anthony Gitter, Rishi Raj Goel, Casey S. Greene, Marouen Ben Guebila, Daniel S. Himmelstein, Fengling Hu, Nafisa M. Jadavji, Jeremy P. Kamil, Sergey Knyazev, Likhitha Kolla, Alexandra J. Lee, Ronan Lordan, Tiago Lubiana, Temitayo Lukan, Adam L. MacLean, David Mai, Serghei Mangul, David Manheim, Lucy D'Agostino McGowan, Amruta Naik, YoSon Park, Dimitri Perrin, Yanjun Qi, Diane N. Rafizadeh, Bharath Ramsundar, Halie M. Rando, Sandipan Ray, Michael P. Robson, Vincent Rubinetti, Elizabeth Sell, Lamonica Shinholster, Ashwin N. Skelly, Yuchen Sun, Yusha Sun, Gregory L Szeto, Ryan Velazquez, Jinhui Wang, Nils Wellhausen

Authors with similar contributions are ordered alphabetically.

## 0.1 Abstract

The novel coronavirus SARS-CoV-2, which emerged in late 2019, has since spread around the world and infected hundreds of millions of people with coronavirus disease 2019 (COVID-19). While this viral species was unknown prior to January 2020, its similarity to other coronaviruses that infect humans has allowed for rapid insight into the mechanisms that it uses to infect human hosts, as well as the ways in which the human immune system can respond. Here, we contextualize SARS-CoV-2 among other coronaviruses and identify what is known and what can be inferred about its behavior once inside a human host. Because the genomic content of coronaviruses, which specifies the virus's structure, is highly conserved, early genomic analysis provided a significant head start in predicting viral pathogenesis and in understanding potential differences among variants. The pathogenesis of the virus offers insights into symptomatology, transmission, and individual susceptibility. Additionally, prior research into interactions between the human immune system and coronaviruses has identified how these viruses can evade the immune system's protective mechanisms. We also explore systems-level research into the regulatory and proteomic effects of SARS-CoV-2 infection and the immune response. Understanding the structure and behavior of the virus serves to contextualize the many facets of the COVID-19 pandemic and can influence efforts to control the virus and treat the disease.

## 0.2 Importance

COVID-19 involves a number of organ systems and can present with a wide range of symptoms. From how the virus infects cells to how it spreads between people, the available research suggests that these patterns are very similar to those seen in the closely related viruses SARS-CoV-1 and possibly MERS-CoV. Understanding the pathogenesis of the SARS-CoV-2 virus also contextualizes how the different biological systems affected by COVID-19 connect. Exploring the structure, phylogeny, and pathogenesis of the virus therefore helps to guide interpretation of the broader impacts of the virus on the human body and on human populations. For this reason, an in-depth exploration of viral mechanisms is critical to a robust understanding of SARS-CoV-2 and, potentially, future emergent HCoV.

## 0.3 Introduction

The current coronavirus disease 2019 (COVID-19) pandemic, caused by the *Severe acute respiratory syndrome-related coronavirus 2* (SARS-CoV-2) virus, represents an acute global health crisis. Symptoms of the disease can range from mild to severe or fatal [1] and can affect a variety of organs and systems [2]. Outcomes of infection can include acute respiratory distress (ARDS) and acute lung injury, as well as damage to other organ systems [2,3]. Understanding the progression of the disease, including these diverse symptoms, depends on understanding how the virus interacts with the host. Additionally, the fundamental biology of the virus can provide insights into how it is transmitted among people, which can, in turn, inform efforts to control its spread. As a result, a thorough understanding of the pathogenesis of SARS-CoV-2 is a critical foundation on which to build an understanding of COVID-19 and the pandemic as a whole.

The rapid identification and release of the genomic sequence of the virus in January 2020 [4] provided early insight into the virus in a comparative genomic context. The viral genomic sequence clusters with known coronaviruses (order *Nidovirales*, family *Coronaviridae*, subfamily *Orthocoronavirinae*). Phylogenetic analysis of the coronaviruses reveals four major subclades, each corresponding to a genus: the alpha, beta, gamma, and delta coronaviruses. Among them, alpha and beta coronaviruses infect mammalian species, gamma coronaviruses infect avian species, and delta coronaviruses infect both mammalian and avian species [5]. The novel virus now known as SARS-CoV-2 was identified as a beta coronavirus belonging to the B lineage based on phylogenetic analysis of a polymerase chain reaction (PCR) amplicon fragment from five patients along with the full genomic sequence [6]. This lineage also includes the *Severe acute respiratory syndrome-related coronavirus* (SARS-CoV-1) that caused the 2002-2003 outbreak of Severe Acute Respiratory Syndrome (SARS) in humans [6]. (Note that these subclades are not to be confused with variants of concern within SARS-CoV-2 labeled with Greek letters; i.e., the Delta variant of SARS-CoV-2 is still a beta coronavirus.)

Because viral structure and mechanisms of pathogenicity are highly conserved within the order, this phylogenetic analysis provided a basis for forming hypotheses about how the virus interacts with hosts, including which tissues, organs, and systems would be most susceptible to SARS-CoV-2 infection. Coronaviruses that infect humans (HCoV) are not common, but prior research into other HCoV such as SARS-CoV-1 and *Middle East respiratory syndrome-related coronavirus* (MERS-CoV), as well as other viruses infecting humans such as a variety of influenza species, established a strong foundation that accelerated the pace of SARS-CoV-2 research.

Coronaviruses are large viruses that can be identified by their distinctive "crown-like" shape (Figure [1]). Their spherical virions are made from lipid envelopes ranging from 100 to 160 nanometers in which peplomers (protruding structures) of two to three spike (S) glycoproteins are anchored, creating the crown [7,8]. These spikes, which are critical to both viral pathogenesis and to the response by the host immune response, have been visualized using cryo-electron microscopy [9]. Because they induce the human immune response, they are also the target of many proposed therapeutic agents [10,11]. Viral pathogenesis is typically broken down into three major components: entry, replication, and spread [12]. However, in order to draw a more complete picture of pathogenesis, it is also necessary to examine how infection manifests clinically, identify systems-level interactions between the virus and the human body, and consider the possible effects of variation or evolutionary change on pathogenesis and virulence. Thus, clinical medicine and traditional biology are both important pieces of the puzzle of SARS-CoV-2 presentation and pathogenesis.

# 0.4  Coronavirus Structure and Pathogenesis

## 0.4.1 Structure of Coronaviruses

Genome structure is highly conserved among coronaviruses, meaning that the relationship between the SARS-CoV-2 genome and its pathogenesis can be inferred from prior research in related viral species. The genomes of viruses in the *Nidovirales* order share several fundamental characteristics. They are non-segmented, which means the viral genome is a single continuous strand of RNA, and are enveloped, which means that the genome and capsid are encased by a lipid bilayer. Coronaviruses have large positive-sense RNA (ssRNA+) genomes ranging from 27 to 32 kilobases in length [13,14]. The SARS-CoV-2 genome lies in the middle of this range at 29,903 bp [14]. Genome organization is highly conserved within the order [13]. There are three major genomic regions: one containing the replicase gene, one containing the genes encoding structural proteins, and interspersed accessory genes [13] (Figure [1]). The replicase gene comprises about two-thirds of the genome and consists of two open reading frames that are translated with ribosomal frameshifting [13]. This polypeptide is then translated into 16 non-structural proteins (nsp), except in gammacoronaviruses where nsp1 is absent, that form the replication machinery used to synthesize viral RNA [15]. The remaining third of the genome encodes structural proteins, including the spike (S), membrane, envelope, and nucleocapsid proteins. Additional accessory genes are sometimes present between these two regions, depending on the species or strain. Much attention has been focused on the S protein, which is a critical structure involved in cell entry.

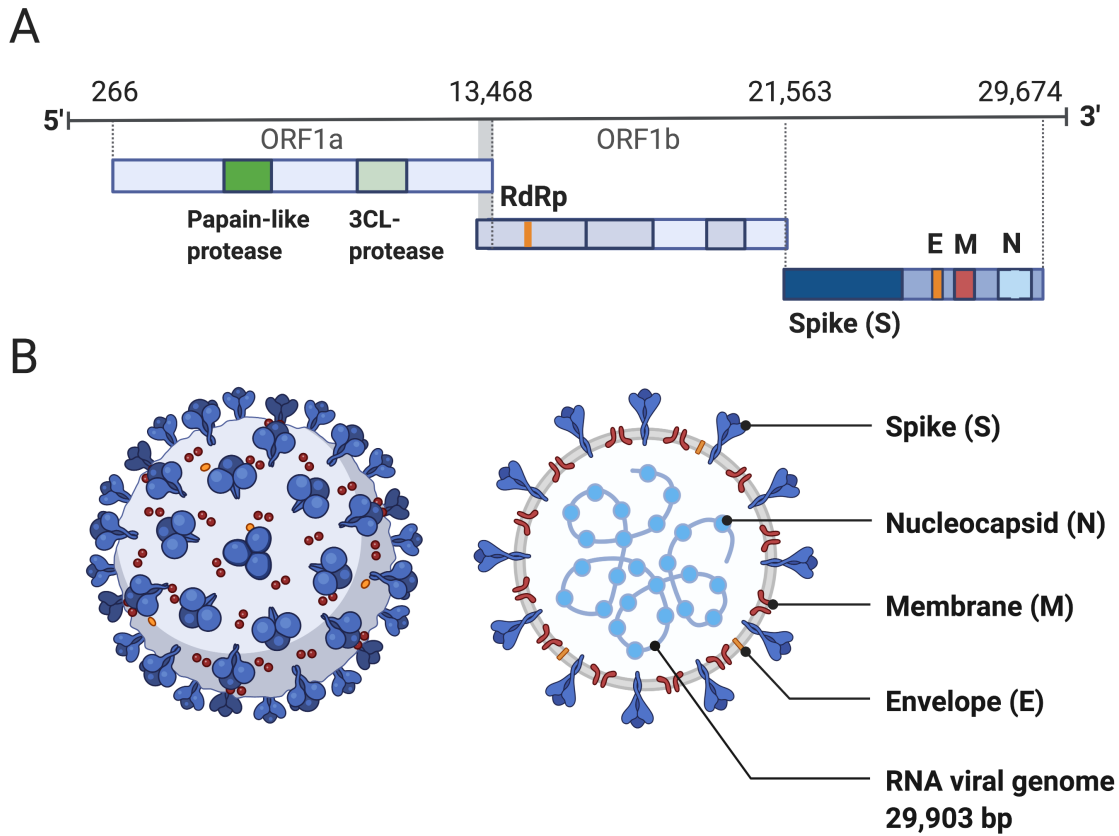

Figure 1: **Structure of SARS-CoV-2 capsid and genome.** A) The genomic structure of coronaviruses is highly conserved and includes three main regions. Open reading frames (ORF) 1a and 1b contain two polyproteins that encode the non-structural proteins (nsp). The nsp include enzymes such as RNA-dependent RNA Polymerase (RdRp). The last third of the genome encodes structural proteins, including the spike (S), envelope (E), membrane (M) and nucleocapsid (N) proteins. Accessory genes can also be interspersed throughout the genome [13]. B) The physical structure of the coronavirus virion, including the components determined by the conserved structural proteins S, E, M and N. This figure was adapted from "Human Coronavirus Structure", by BioRender.com (2020), retrieved from https://app.biorender.com/biorender-templates.

## 0.4.2 Pathogenic Mechanisms of Coronaviruses

While it is possible that SARS-CoV-1 and SARS-CoV-2, like most viruses, enter cells through endocytosis, a process conserved among coronaviruses enables them to target cells for entry through fusion with the plasma membrane [16,17]. Cell entry proceeds in three steps: binding, cleavage, and fusion. First, the viral spike protein binds to a host cell via a recognized receptor or entry point. Coronaviruses can bind to a range of host receptors [18,19], with binding conserved only at the genus level [5]. Viruses in the beta coronavirus genus, to which SARS-

CoV-2 belongs, are known to bind to the CEACAM1 protein, 5-N-acetyl-9-O-acetyl neuraminic acid, and to angiotensin-converting enzyme 2 (ACE2) [18]. This recognition is driven by domains in the S1 subunit [20]. SARS-CoV-2 has a high affinity for human ACE2, which is expressed in the vascular epithelium, other epithelial cells, and cardiovascular and renal tissues [21,22], as well as many others [23]. The binding process is guided by the molecular structure of the spike protein, which is structured in three segments: an ectodomain, a transmembrane anchor, and an intracellular tail [24]. The ectodomain forms the crown-like structures on the viral membrane and contains two subdomains known as the S1 and S2 subunits [25]. The S1 (N-terminal) domain forms the head of the crown and contains the receptor binding motif, and the S2 (C-terminal) domain forms the stalk that supports the head [25]. The S1 subunit guides the binding of the virus to the host cell, and the S2 subunit guides the fusion process [24].

After the binding of the S1 subunit to an entry point, the spike protein of coronaviruses is often cleaved at the S1/S2 boundary into the S1 and S2 subunits by a host protease [20,26,27]. This proteolytic priming is important because it prepares the S protein for fusion [26,27]. The two subunits remain bound by van der Waals forces, with the S1 subunit stabilizing the S2 subunit throughout the membrane fusion process [20]. Cleavage at a second site within S2 (S2') activates *S* for fusion by inducing conformational changes [20]. Similar to SARS-CoV-1, SARS-CoV-2 exhibits redundancy in which host proteases can cleave the S protein [28]. Both transmembrane protease serine protease-2 (TMPRSS-2) and cathepsins B/L have been shown to mediate SARS-CoV-2 S protein proteolytic priming, and small molecule inhibition of these enzymes fully inhibited viral entry *in vitro* [28,29]. Other proteases known to cleave the S1/S2 boundary in coronaviruses include TMPRSS-4, trypsin, furin, cathepsins, and human airway trypsin-like protease (HAT) [29].

Unlike in SARS-CoV-1, a second cleavage site featuring a furin-like binding motif is also present near the S1/S2 boundary in SARS-CoV-2 [30]. This site is found in HCoV belonging to the A and C lineages of beta coronavirus, including MERS-CoV, but not in the other known members of the B lineage of beta coronavirus that contains SARS-CoV-1 and SARS-CoV-2 [30]. It is associated with increased virulence in other viral species [30] and may facilitate membrane fusion of SARS-CoV-2 in the absence of other proteases that prime the S1/S2 site [31]. However, given that proteases such as HAT are likely to be present in targets like the human airway, the extent to which this site has had a real-world effect on the spread of SARS-CoV-2 was initially unclear [31]. Subsequent research has supported this site as an important contributor to pathogenesis: *in vitro* analyses have reported that it bolsters pathogenicity specifically in cell lines derived from human airway cells (Calu3 cell line) [32,33,34] and that furin inhibitors reduced pathogenic effects in VeroE6 cells [35].

Electron microscopy suggests that in some coronaviruses, including SARS-CoV-1 and MERS-CoV, a six-helix bundle separates the two subunits in the postfusion conformation, and the unusual length of this bundle facilitates membrane fusion through the release of additional energy [5]. The viral membrane can then fuse with the endosomal membrane to release the viral genome into the host cytoplasm. Once the virus enters a host cell, the replicase gene is translated and assembled into the viral replicase complex. This complex then synthesizes the double-stranded RNA (dsRNA) genome from the genomic ssRNA(+). The dsRNA genome is transcribed and replicated to create viral mRNAs and new ssRNA(+) genomes [13,36]. From there, the virus can spread into other cells. In SARS-CoV-2, the insertion of the furin-like binding site near the S1/S2 boundary is also thought to increase cell-cell adhesion, making it possible for the viral genome to spread directly from cell to cell rather than needing to propagate the virion itself [37]. In this way, the genome of SARS-CoV-2 provides insight into the pathogenic behavior of the virus.

Evidence also suggests that SARS-CoV-2 may take advantage of the specific structure of endothelial cells to enter the circulatory system. Endothelial cells are specialized epithelial cells [38] that form a barrier between the bloodstream and surrounding tissues. The endothelium facilitates nutrient, oxygen, and cellular exchange between the blood and vascularized tissues [39]. The luminal (interior) surface of the endothelium is lined with glycocalyx, a network of both membrane-bound and soluble proteins and carbohydrates, primarily proteoglycans and glycoproteins [40,41]. The glycocalyx varies in thickness from 0.5 microns in the capillaries to 4.5 microns in the carotid arteries and forms a meshwork that localizes both endothelial- and plasma-derived signals to the inner vessel wall [40]. Heparan sulfate is the dominant proteoglycan in the glycocalyx, representing 50-90% of glycocalyx proteoglycan content [42]. The SARS-CoV-2 spike protein can bind directly to heparan sulfate, which serves in part as a scaffolding molecule to facilitate ACE2 binding and entry into endothelial cells [41]. A heparan sulfate binding site has also been identified near the ACE2 binding site on the viral receptor binding domain (RBD), and modeling has suggested that heparan sulfate binding yields an open conformation that facilitates binding to ACE2 on the cell surface [41]. Degrading or removing heparan sulfate was associated with decreased binding [41]. Heparan sulfate may also interact with the S1/S2 proteolytic cleavage site and other binding sites to promote binding affinity [43]. Notably, treatment with soluble heparan sulfate or even heparin (a commonly used anti-coagulant and vasodilator that is similar in structure to heparan sulfate [44]) potently blocked spike protein binding and viral infection [41]. This finding is particularly interesting because degradation of heparan sulfate in the glycocalyx has previously been identified as an important contributor to ARDS and sepsis [45], two common and severe outcomes of COVID-19, and suggests that heparan sulfate could be a target for pharmaceutical inhibition of cell entry by SARS-CoV-2 [46,47,48,49,50]. Together, this evidence suggests that heparan sulfate can serve as an important adhesion molecule for SARS-CoV-2 cell entry. It may represent a therapeutic target but has not been pursued as much as other candidate targets [10].

## 0.4.3 Immune Evasion Strategies

Research in other HCoV provides some indication of how SARS-CoV-2 infection can proceed despite human immune defenses. Infecting the epithelium can help viruses such as SARS-CoV-1 bypass the physical barriers, such as mucus, that comprise the immune system's first line of defense [51]. Once the virus infiltrates host cells, it is adept at evading detection. CD163+ and CD68+ macrophage cells are especially crucial for the establishment of SARS-CoV-1 in the body [51]. These cells most likely serve as viral reservoirs that help shield SARS-CoV-1 from the innate immune response. According to a study on the viral dissemination of SARS-CoV-1 in Chinese macaques, viral RNA could be detected in some monocytes throughout the process of differentiation into dendritic cells [51]. This lack of active viral replication allows SARS-CoV-1 to escape the innate immune response because reduced levels of detectable viral RNA allow the virus to avoid both natural killer cells and Toll-like receptors [51]. Even during replication, SARS-CoV-1 is able to mask its dsRNA genome from detection by the immune system. Although dsRNA is a pathogen-associated molecular pattern that would typically initiate a response from the innate immune system [52], *in vitro* analysis of nidoviruses including SARS-CoV-1 suggests that these viruses can induce the development of double-membrane vesicles that protect the dsRNA signature from being detected by the host immune system [53]. This protective envelope can therefore insulate these coronaviruses from the innate immune system's detection mechanism [54].

HCoVs are also known to interfere with the host immune response, rather than just evade it. For example, the virulence of SARS-CoV-2 is increased by nsp1, which can suppress host gene expression by stalling mRNA translation and inducing endonucleolytic cleavage and mRNA degradation [55]. SARS-CoV-1 also evades the immune response by interfering with type I IFN induction signaling, which is a mechanism that leads to cellular resistance to viral infections. SARS-CoV-1 employs methods such as ubiquitination and degradation of RNA sensor adaptor molecules MAVS and TRAF3/6 [56]. Also, MERS-CoV downregulates antigen presentation via MHC class I and MHC class II, which leads to a reduction in T cell activation [56]. These evasion mechanisms, in turn, may facilitate systemic infection. Coronaviruses such as SARS-CoV-1 are also able to evade the humoral immune response through other mechanisms, such as inhibiting certain cytokine pathways or down-regulating antigen presentation by the cells [53].

## 0.4.4 Host Cell Susceptibility

ACE2 and TMPRSS-2 have been identified as the primary entry portal and as a critical protease, respectively, in facilitating the entry of SARS-CoV-1 and SARS-CoV-2 into a target cell [9,28,57,58,59]. This finding has led to a hypothesized role for the expression of these molecules in determining which cells, tissues, and organs are most susceptible to SARS-CoV-2 infection. ACE2 is expressed in numerous organs, such as the heart, kidney, and intestine, but it is most prominently expressed in alveolar epithelial cells; this pattern of expression is expected to contribute to the virus' association with lung pathology [21,60,61] as well as that of SARS [62]. A retrospective observational study reported indirect evidence that certain antineoplastic therapies, such as the chemotherapy drug gemcitabine, may reduce risk of SARS-CoV-2 infection in patients with cancer, possibly via decreased ACE2 expression [63]. Additionally, the addition of the furin site insertion at the S1/S2 boundary means that SARS-CoV-2 does not require TMPRSS-2 when furin, an ubiquitously expressed endoprotease [64], is present, enabling cell-cell fusion independent of TMPRSS-2 availability [65].

Clinical investigations of COVID-19 patients have detected SARS-CoV-2 transcripts in bronchoalveolar lavage fluid (BALF) (93% of specimens), sputum (72%), nasal swabs (63%), fibrobronchoscopy brush biopsies (46%), pharyngeal swabs (32%), feces (29%), and blood (1%) [66]. Two studies reported that SARS-CoV-2 could not be detected in urine specimens [66,67]; however, a third study identified four urine samples (out of 58) that were positive for SARS-CoV-2 nucleic acids [68]. Although respiratory failure remains the leading cause of death for COVID-19 patients [69], SARS-CoV-2 infection can damage many other organ systems including the heart [70], kidneys [71,72], liver [73], and gastrointestinal tract [74,75]. As it becomes clear that SARS-CoV-2 infection can damage multiple organs, the scientific community is pursuing multiple avenues of investigation in order to build a consensus about how the virus affects the human body.

## 0.5  Clinical Presentation of COVID-19

SARS-CoV-2 pathogenesis is closely linked with the clinical presentation of the COVID-19 disease. Reports have described diverse symptom profiles associated with COVID-19, with a great deal of variability both within and between institutions and regions. Definitions for non-severe, severe, and critical COVID-19, along with treatment recommendations, are available from the World Health Organization living guidelines [76]. A large study from Wuhan, China conducted early in the pandemic identified fever and cough as the two most common symptoms that patients reported at hospital admission [77], while a retrospective study in China described the clinical presentations of patients infected with SARS-CoV-2 as including lower respiratory tract infection with fever, dry cough, and dyspnea (shortness of breath) [78]. This study [78] noted that upper respiratory tract symptoms were less common, suggesting that the virus preferentially targets cells located in the lower respiratory tract. However, data from the New York City region [79,80] showed variable rates of fever as a presenting symptom, suggesting that symptoms may not be consistent across individuals. For example, even within New York City, one study [79] identified low oxygen saturation (<90% without the use of supplemental oxygen or ventilation support) in 20.4% of patients upon presentation, with fever being present in 30.7%, while another study [80] reported cough (79.4%), fever (77.1%), and dyspnea (56.5%) as the most common presenting symptoms; both of these studies considered only hospitalized patients. A later study reported radiographic findings such as ground-glass opacity and bilateral patchy shadowing in the lungs of many hospitalized patients, with most COVID-19 patients having lymphocytopenia, or low levels of lymphocytes (a type of white blood cell) [77]. Patients may also experience loss of smell, myalgias (muscle aches), fatigue, or headache. Gastrointestinal symptoms can also present [81], and the CDC includes nausea and vomiting, as well as congestion and runny nose, on its list of symptoms consistent with COVID-19 [1]. An analysis of an app-based survey of 500,000 individuals in the U.S. found that among those tested for SARS-CoV-2, a loss of taste or smell, fever, and a cough were significant predictors of a positive test result [82]. It is important to note that in this study, the predictive value of symptoms may be underestimated if they are not specific to COVID-19. This underestimation could occur because the outcome measured was a positive, as opposed to a negative, COVID-19 test result, meaning an association would be more easily identified for symptoms that were primarily or exclusively found with COVID-19. At the time the surveys were conducted, due to limits in U.S. testing infrastructure, respondents typically needed to have some symptoms known to be specific to COVID-19 in order to qualify for testing. Widespread testing of asymptomatic individuals may therefore provide additional insight into the range of symptoms associated with COVID-19.

Consistent with the wide range of symptoms observed and the pathogenic mechanisms described above, COVID-19 can affect a variety of systems within the body in addition to causing respiratory problems [83]. For example, COVID-19 can lead to acute kidney injury, especially in patients with severe respiratory symptoms or certain preexisting conditions [84]. Some patients are at risk for collapsing glomerulopathy [85].

COVID-19 can also cause neurological complications [86,87,88], potentially including stroke, seizures or meningitis [89,90]. One study on autopsy samples suggested that SARS-CoV-2 may be able to enter the central nervous system via the neural–mucosal interface [91]. However, a study of 41 autopsied brains [92] found no evidence that the virus can actually infect the central nervous system. Although there was viral RNA in some brain samples, it was only found in very small amounts, and no viral protein was found. The RNA may have been in the blood vessels or blood components and not in the brain tissue itself. Instead, the neuropathological effects of COVID-19 are more likely to be caused indirectly by hypoxia, coagulopathy, or inflammatory processes rather than by infection in the brain [92]. COVID-19 has been associated with an increased incidence of large vessel stroke, particularly in patients under the age of 40 [93], and other thrombotic events including pulmonary embolism and deep vein thrombosis [94]. The mechanism behind these complications has been suggested to be related to coagulopathy, with reports indicating the presence of antiphospholipid antibodies [95] and elevated levels of d-dimer and fibrinogen degradation products in deceased patients [96]. Other viral infections have been associated with coagulation defects and changes to the coagulation cascade; notably, SARS was also found to lead to disseminated intravascular coagulation and was associated with both pulmonary embolism and deep vein thrombosis [97]. The mechanism behind these insults has been suggested to be related to inflammation-induced increases in the von Willebrand factor clotting protein, leading to a pro-coagulative state [97]. Abnormal clotting (thromboinflammation or coagulopathy) has been increasingly discussed recently as a possible key mechanism in many cases of severe COVID-19, and may be associated with the high d-dimer levels often observed in severe cases [98,99,100]. This excessive clotting in lung capillaries has been suggested to be related to a dysregulated activation of the complement system, part of the innate immune system [101,102].

Finally, concerns have been raised about long-term sequelae of COVID-19. Some COVID-19 patients have reported that various somatic symptoms (such as shortness of breath, fatigue, chest pain) and psychological (depression, anxiety or mild cognitive impairment) symptoms can last for months after infection [103]. Such long-term affects occur both in adults [104] and children [105]. Sustained symptoms affecting a variety of biological systems have been reported across many studies (e.g., [103,106,107]). The phenomenon of "long COVID" is not fully understood although various possible explanations have been proposed, including damage caused by immune response to infection as well as by the infection itself, in addition to negative consequences of the experience of lengthy illness and hospitalization. However, a lack of consistency among definitions used in different studies makes it difficult to develop precise definitions or identify specific symptoms associated with long-term effects of COVID-19 [108,109]. Patient and family support groups for "long haulers" have been formed online, and patient-driven efforts to collect data about post-acute COVID-19 provide valuable sources of information (e.g., [106]). The specific relationship between viral pathogenesis and these reported sequelae remains to be uncovered, however.

## 0.5.1 Pediatric Presentation

The presentation of COVID-19 infection can vary greatly among pediatric patients and, in some cases, manifests in distinct ways from COVID-19 in adults. Evidence suggests that children and adolescents tend to have mostly asymptomatic infections and that those who are symptomatic typically exhibit mild illness [110,111,112,113]. One review examined symptoms reported in 17 studies of children infected with COVID-19 during the early months of the COVID-19 epidemic in China and one study from Singapore [114]. In the more than a thousand cases described, the most common reports were for mild symptoms such as fever, dry cough, fatigue, nasal congestion and/or runny nose, while three children were reported to be asymptomatic. Severe lower respiratory infection was described in only one of the pediatric cases reviewed. Gastrointestinal symptoms such as vomiting or diarrhea were occasionally reported. Radiologic findings were not always reported in the case studies reviewed, but when they were mentioned they included bronchial thickening, ground-glass opacities, and/or inflammatory lesions [114]. Neurological symptoms have also been reported [115].

These analyses indicate that most pediatric cases of COVID-19 are not severe. Indeed, it is estimated that less than 1% of pediatric cases result in critical illness [112,116], although reporting suggests that pediatric hospitalizations may be greater with the emergence of the Delta variant of concern (VOC) [117,118,119]. Serious complications and, in relatively rare cases, deaths have occurred [120]. Of particular interest, children have occasionally experienced a serious inflammatory syndrome, multisystem inflammatory syndrome in children (MIS-C), following COVID-19 infection [121]. This syndrome is similar in some respects to Kawasaki disease, including Kawasaki disease shock syndrome [122,123,124], and is thought to be a distinct clinical manifestation of SARS-CoV-2 due to its distinct cytokine profile and the presence of burr cells in peripheral blood smears [125,126]. MIS-C has been associated with heart failure in some cases [127]. A small number of case studies have identified presentations similar to MIS-C in adults associated with SARS-CoV-2 [128,129,130,131]. However, not all cases of severe COVID-19 in children are characterizable as MIS-C. A recent study [132] described demographic and clinical variables associated with MIS-C in comparison with non-MIS-C severe acute COVID-19 in young people in the United States. Efforts to characterize long-term sequelae of SARS-CoV-2 infection in children face the same challenges as in adults, but long-term effects remain a concern in pediatric patients [105,133,134], although some early studies have suggested that they may be less of a concern than in adults [135,136,137]. Research is ongoing into the differences between the pediatric and adult immune responses to SARS-CoV-2, and future research may shed light on the factors that lead to MIS-C; it is also unknown whether the relative advantages of children against severe COVID-19 will remain in the face of current and future variants [138].

## 0.5.2 Cytokine Release Syndrome

The inflammatory response was identified early on as a potential driver of COVID-19 outcomes due to existing research in SARS and emerging research in COVID-19. While too low of an inflammatory response is a concern because it will fail to eliminate the immune threat [139], excessive pro-inflammatory cytokine activity can cascade [140] and cause cell damage, among other problems [141]. A dysregulated immune response can cause significant damage to the host [142,143,144] including pathogenesis associated with sepsis. Sepsis, which can lead to multi-organ failure and death [145,146], is traditionally associated with bacterial infections. However, sepsis associated with viral infections may be underidentified [147], and sepsis has emerged as a major concern associated with SARS-CoV-2 infection [148]. Hyperactivity of the pro-inflammatory response due to lung infection is commonly associated with acute lung injury and more rarely with the more severe manifestation, ARDS, which can arise from pneumonia, SARS, and COVID-19 [140,145]. Damage to the capillary endothelium can cause leaks that disrupt the balance between pro-inflammatory cytokines and their regulators [149], and heightened inflammation in the lungs can also serve as a source for systemic inflammation, or sepsis, and potentially multi-organ failure [145]. The shift from local to systemic inflammation is a phenomenon often referred to broadly as a cytokine storm [145] or, more precisely, as cytokine release syndrome [150].

Cytokine dysregulation is therefore a significant concern in the context of COVID-19. In addition to the known role of cytokines in ARDS and lung infection more broadly, immunohistological analysis at autopsy of deceased SARS patients revealed that ACE2-expressing cells that were infected by SARS-CoV-1 showed elevated expression of the cytokines IL-6, IL-1β, and TNF-α [151]. Similarly, the introduction of the S protein from SARS-CoV-1 to mouse macrophages was found to increase production of IL-6 and TNF-α [152]. For SARS-CoV-2 infection leading to COVID-19, early reports described a cytokine storm syndrome-like response in patients with particularly severe infections [60,153,154]. Sepsis has been identified as a major contributor to COVID-19-related death. Among patients hospitalized with COVID-19 in Wuhan, China, 112 out of 191 (59%) developed sepsis, including all 54 of the non-survivors [78].

While IL-6 is sometimes used as a biomarker for cytokine storm activity in sepsis [145], the relationship between cytokine profiles and the risks associated with sepsis may be more complex. One study of patients with and at risk for ARDS, specifically those who were intubated for medical ventilation, found that shortly after the onset of ARDS, anti-inflammatory cytokine concentration in BALF increased relative to the concentration of pro-inflammatory cytokines [149]. The results suggest that an increase in pro-inflammatory cytokines such as IL-6 may signal the onset of ARDS, but recovery depends on an increased anti-inflammatory response [149]. However, patients with severe ARDS were excluded from this study. Another analysis of over 1,400 pneumonia patients in the United States reported that IL-6, tumor necrosis factor (TNF), and IL-10 were elevated at intake in patients who developed severe sepsis and/or ultimately died [155]. However, unlike the study analyzing pro- and anti-inflammatory cytokines in ARDS patients [149], this study reported that unbalanced pro-/anti-inflammatory cytokine profiles were rare. This discrepancy could be related to the fact that the sepsis study measured only three cytokines. Although IL-6 has traditionally been considered pro-inflammatory, its pleiotropic effects via both classical and trans-signaling allow it to play an integral role in both the inflammatory and anti-inflammatory responses [156], leading it to be associated with both healthy and pathological responses to viral threat [157]. While the cytokine levels observed in COVID-19 patients fall outside of the normal range, they are not as high as typically found in patients with ARDS [158]. Regardless of variation in the anti-inflammatory response, prior work has therefore made it clear that pulmonary infection and injury are associated with systemic inflammation and with sepsis. Inflammation has received significant interest both in regards to the pathology of COVID-19 as well as potential avenues for treatment, as the relationship between the cytokine storm and the pathophysiology of COVID-19 has led to the suggestion that a number of immunomodulatory pharmaceutical interventions could hold therapeutic value for the treatment of COVID-19 [10,159].

## 0.6  Insights from Systems Biology

Systems biology provides a cross-disciplinary analytical paradigm through which the host response to an infection can be analyzed. This field integrates the "omics" fields (genomics, transcriptomics, proteomics, metabolomics, etc.) using bioinformatics and other computational approaches. Over the last decade, systems biology approaches have been used widely to study the pathogenesis of diverse types of life-threatening acute and chronic infectious diseases [160]. Omics-based studies have also provided meaningful information regarding host immune responses and surrogate protein markers in several viral, bacterial and protozoan infections [161]. Though the complex pathogenesis and clinical manifestations of SARS-CoV-2 infection are not yet fully understood, omics technologies offer the opportunity for discovery-driven analysis of biological changes associated with SARS-CoV-2 infection.

## 0.6.1 Transcriptomics

Through transcriptomic analysis, the effect of a viral infection on gene expression can be assessed. Transcriptomic analyses, whether *in vivo* or *in situ*, can potentially reveal insights into viral pathogenesis by elucidating the host response to the virus. For example, infection by some viruses, including by the coronaviruses SARS-CoV-2, SARS-CoV-1, and MERS-CoV, is associated with the upregulation of ACE2 in human embryonic kidney cells and human airway epithelial cells [60]. This finding suggests that SARS-CoV-2 facilitates the positive regulation of its own transmission between host cells [60]. The host immune response also likely plays a key role in mediating infection-associated pathologies. Therefore, transcriptomics is one critical tool for characterizing the host response in order to gain insight into viral pathogenesis. For this reason, the application of omics technologies to the process of characterizing the host response is expected to provide novel insights into how hosts respond to SARS-CoV-2 infection and how these changes might influence COVID-19 outcomes.

Several studies have examined the cellular response to SARS-CoV-2 *in vitro* in comparison to other viruses. One study [162] compared the transcriptional responses of three human cell lines to SARS-CoV-2 and to other respiratory viruses, including MERS-CoV, SARS-CoV-1, *Human parainfluenza virus 3*, *Respiratory syncytial virus*, and *Influenza A virus*. The transcriptional response differed between the SARS-CoV-1 infected cells and the cells infected by other viruses, with changes in differential expression specific to each infection type. Where SARS-CoV-2 was able to replicate efficiently, differential expression analysis revealed that the transcriptional response was significantly different from the response to all of the other viruses tested. A unique pro-inflammatory cytokine signature associated with SARS-CoV-2 was present in cells exposed to both high and low doses of the virus, with the cytokines IL-6 and IL1RA uniquely elevated in response to SARS-CoV-2 relative to other viruses. However, one cell line showed significant IFN-I or IFN-III expression when exposed to high, but not low, doses of SARS-CoV-2, suggesting that IFN induction is dependent on the extent of exposure. These results suggest that SARS-CoV-2 induces a limited antiviral state with low IFN-I or IFN-III expression and a moderate IFN-stimulated gene response, in contrast to other viruses. Other respiratory viruses have been found to encode antagonists to the IFN response [163,164], including SARS-CoV-1 [165] and MERS-CoV [166].

The analysis of SARS-CoV-2 suggested that this transcriptional state was specific to cells expressing ACE2, as it was not observed in cells lacking expression of this protein except with ACE2 supplementation and at very high (10-fold increase) level of SARS-CoV-2 exposure [162]. In another study, direct stimulation with inflammatory cytokines such as type I interferons (e.g.,

IFNβ) was also associated with the upregulation of ACE2 in human bronchial epithelial cells, with treated groups showing four-fold higher ACE2 expression than control groups at 18 hours post-treatment [167]. This hypothesis was further supported by studies showing that several nsp in SARS-CoV-2 suppress interferon activity [168] and that the SARS-CoV-2 *ORF3b* gene suppresses IFNB1 promoter activity (IFN-I induction) more efficiently than the SARS-CoV-1 *ORF3b* gene [169]. Taken together, these findings suggest that a unique cytokine profile is associated with the response to the SARS-CoV-2 virus, and that this response differs depending on the magnitude of exposure.

Susceptibility and IFN induction may also vary by cell type. Using poly(A) bulk RNA-seq to analyzed dynamic transcriptional responses to SARS-CoV-2 and SARS-CoV-1 revealed negligible susceptibility of cells from the H1299 line (< 0.08 viral read percentage of total reads) compared to those from the Caco-2 and Calu-3 lines (>10% of viral reads) [170]. This finding suggests that the risk of infection varies among cell types, and that cell type could influence which hosts are more or less susceptible. Based on visual inspection of microscopy images alongside transcriptional profiling, the authors also showed distinct responses among the host cell lines evaluated [170]. In contrast to Caco-2, Calu-3 cells infected with SARS-CoV-2 showed signs of impaired growth and cell death at 24 hours post infection, as well as moderate IFN induction with a strong up-regulation of IFN-stimulated genes. Interestingly, the results were similar to those reported in Calu-3 cells exposed to much higher levels of SARS-CoV-2 [162], as described above. This finding suggests that IFN induction in Calu-3 cells is not dependent on the level of exposure, in contrast to A549-ACE2 cells. The discrepancy could be explained by the observations that Calu-3 cells are highly susceptible to SARS-CoV-2 and show rapid viral replication [29], whereas A549 cells are incompatible with SARS-CoV-2 infection [171]. This discrepancy raises the concern that *in vitro* models may vary in their similarity to the human response, underscoring the importance of follow-up studies in additional models.

As a result, transcriptional analysis of patient tissue is an important application of omics technology to understanding COVID-19. Several studies have collected blood samples from COVID-19 patients and analyzed them using RNA-Seq [172,173,174,175,176,177]. Analyzing gene expression in the blood is valuable to understanding host-pathogen interactions because of the potential to identify alterations associated with the immune response and to gain insights into inflammation, among other potential insights [172]. One study compared gene expression in 39 COVID-19 inpatients admitted with community-acquired pneumonia to that of control donors using whole blood cell transcriptomes [172]. They also evaluated the effect of mild versus severe disease. A greater number of differentially expressed genes were found in severe patients compared to controls than in mild patients compared to controls. They also identified that the transcriptional profiles clustered into five groups and that the groups could not be explained by disease severity. Most severe cases fell into two clusters associated with increased inflammation and granulocyte and neutrophil activation. The presence of these clusters suggests the possibility that personalized medicine could be useful in the treatment of COVID-19 [172]. Longitudinal analysis of granulocytes from patients with mild versus severe COVID-19 revealed that granulocyte activation-associated factors differentiated the disease states, with greater numbers of differentially expressed genes early in disease course [172]. This study therefore revealed distinct patterns associated with COVID-19 and identified genes and pathways associated with each cluster.

Many other studies have also identified transcriptomic signatures associated with the immune response and inflammation. Other studies have profiled the transcriptome of BALF [174] and the nasopharynx [178]. One study used single-cell transcriptomics techniques to investigate cell types including brain and choroid plexus cells compared to healthy controls and controls with influenza; among other signals of neuroinflammation, this study reported cortical T cells only in COVID-19 patients [179]. Transcriptomic analysis can thus provide insight into the pathogenesis of SARS-CoV-2 and may also be useful in identifying candidate therapeutics [172].

## 0.6.2 Proteomics

Proteomics analysis offers an opportunity to characterize the response to a pathogen at a level above transcriptomics. Especially early on, this primarily involved evaluating the effect of the virus on cell lines. One early proteomics study investigated changes associated with *in vitro* SARS-CoV-2 infection using Caco-2 cells [180]. This study reported that SARS-CoV-2 induced alterations in multiple vital physiological pathways, including translation, splicing, carbon metabolism and nucleic acid metabolism in the host cells. Another area of interest is whether SARS-CoV-2 is likely to induce similar changes to other HCoV. For example, because of the high level of sequence homology between SARS-CoV-2 and SARS-CoV-1, it has been hypothesized that sera from convalescent SARS-CoV-1 patients might show some efficacy in cross-neutralizing SARS-CoV-2-S-driven entry [28]. However, despite the high level of sequence homology, certain protein structures might be immunologically distinct, which would be likely to prohibit effective cross-neutralization across different SARS species [181]. Consequently, proteomic analyses of SARS-CoV-1 might also provide some essential information regarding the new pathogen [182,183].

Proteomics research has been able to get ahead of the timeline for development of omics-level big data sets specific to SARS-CoV-2 by adopting a comparative bioinformatics approach. Data hubs such as UniProt [184], NCBI Genome Database [185], The Immune Epitope Database and Analysis Resource [186], and The Virus Pathogen Resource [187] contain a wealth of data from studies in other viruses and even HCoV. Such databases facilitate the systems-level reconstruction of protein-protein interaction networks, providing opportunities to generate hypotheses about the mechanism of action of SARS-CoV-2 and identify potential drug targets. In an initial study [188], 26 of the 29 SARS-CoV-2 proteins were cloned and expressed in HEK293T kidney cells, allowing for the identification of 332 high-confidence human proteins interacting with them. Notably, this study suggested that SARS-CoV-2 interacts with innate immunity pathways. Ranking pathogens by the similarity between their interactomes and that of SARS-CoV-2 suggested *West Nile virus*, *Mycobacterium tuberculosis*, and *human papillomavirus* infections as the top three hits. The fact that the host-pathogen interactome of the bacterium *Mycobacterium tuberculosis* was found to be similar to that of SARS-CoV-2 suggests that changes related to lung pathology might comprise a significant contributor to these expression profiles. Additionally, it was suggested that the envelope protein, E, could disrupt host bromodomain-containing proteins, i.e., BRD2 and BRD4, that bind to histones, and the spike protein could likely intervene in viral fusion by modulating the GOLGA7-ZDHHC5 acyl-transferase complex to increase palmitoylation, which is a post-translational modification that affects how proteins interact with membranes [189].

An example of an application of this *in silico* approach comes from another study [190], which used patient-derived peripheral blood mononuclear cells to identify 251 host proteins targeted by SARS-CoV-2. This study also reported that more than 200 host proteins were disrupted following infection. In particular, a network analysis showed that nsp9 and nsp10 interacted with NF-Kappa-B-Repressing Factor, which encodes a transcriptional repressor that mediates repression of genes responsive to Nuclear Factor kappa-light-chain-enhancer of activated B-cells. These genes are important to pro-, and potentially also anti-, inflammatory signaling [191]. This finding could explain the exacerbation of the immune response that shapes the pathology and the high cytokine levels characteristic of COVID-19, possibly due to the chemotaxis of neutrophils mediated by IL-8 and IL-6. Finally, it was suggested [192] that the E protein of both SARS-CoV-1 and SARS-CoV-2 has a conserved Bcl-2 Homology 3-like motif, which could inhibit anti-apoptosis proteins, e.g., BCL2, and trigger the apoptosis of T cells. Several compounds are known to disrupt the host-pathogen protein interactome, largely through the inhibition of host proteins. Therefore, this research identifies candidate targets for intervention and suggests that drugs modulating protein-level interactions between virus and host could be relevant to treating COVID-19.

As with other approaches, analyzing the patterns found in infected versus healthy human subjects is also important. COVID-19 infection has been associated with quantitative changes in transcripts, proteins, metabolites, and lipids in patient blood samples [193]. One longitudinal study [194] compared COVID-19 patients to symptomatic controls who were PCR-negative for SARS-CoV-2. The longitudinal nature of this study allowed it to account for differences in the scale of inter- versus intraindividual changes. At the time of first sampling, common functions of proteins upregulated in COVID-19 patients relative to controls were related to immune system mediation, coagulation, lipid homeostasis, and protease inhibition. They compared these data to the patient-specific timepoints associated with the highest levels of SARS-CoV-2 antibodies and found that the actin-binding protein gelsolin, which is involved in recovery from disease, showed the steepest decline between those two timepoints. Immunoglobulins comprised the only proteins that were significantly different between the COVID-19 and control patients at both of these timepoints. The most significantly downregulated proteins between these time points were related to inflammation, while the most significantly upregulated proteins were immunoglobulins. Proteins related to coagulation also increased between the two timepoints. The selection of a symptomatic control cohort rather than healthy comparisons also suggests that the results are more likely to highlight the response to SARS-CoV-2 and COVID-19 specifically, rather than to disease more broadly. This study also compared the disease course in patients who ultimately survived to those who died and found that ITIH4, a protein associated with the inflammatory response to trauma, may be a biomarker useful to identifying patients at risk of death. Thus, these results indicate the value of studying patients in a longitudinal manner over the disease course. By revealing which genes are perturbed during SARS-CoV-2 infection, proteomics-based analyses can thus provide novel insights into host-virus interaction and serve to generate new avenues of investigation for therapeutics.

## 0.7 Viral Virulence

Like that of SARS-CoV-1, the entry of SARS-CoV-2 into host cells is mediated by interactions between the viral spike glycoprotein, S, and human ACE2 (hACE2)

[20,28,195,196,197,198,199,200]. Differences in how the S proteins of the two viruses interact with hACE2 could partially account for the increased transmissibility of SARS-CoV-2. Studies have reported conflicting binding constants for the S-hACE2 interaction, though they have agreed that the SARS-CoV-2 S protein binds with equal, if not greater, affinity than the SARS-CoV-1 S protein does [9,20,198]. The C-terminal domain of the SARS-CoV-2 S protein in particular was identified as the key region of the virus that interacts with hACE2, and the crystal structure of the C-terminal domain of the SARS-CoV-2 S protein in complex with hACE2 reveals stronger interaction and a higher affinity for receptor binding than that of SARS-CoV-1 [199]. Among the 14 key binding residues identified in the SARS-CoV-1 S protein, eight are conserved in SARS-CoV-2, and the remaining six are semi-conservatively substituted, potentially explaining variation in binding affinity [20,198]. Studies of crystal structure have shown that the RBD of the SARS-CoV-2 S protein, like that of other coronaviruses, undergoes stochastic hinge-like movement that flips it from a "closed" conformation, in which key binding residues are hidden at the interface between protomers, to an "open" one [9,20]. Spike proteins cleaved at the furin-like binding site are substantially more likely to take an open conformation (66%) than those that are uncleaved (17%) [201]. Because the RBD plays such a critical role in viral entry, blocking its interaction with ACE2 could represent a promising therapeutic approach. Nevertheless, despite the high structural homology between the SARS-CoV-2 RBD and that of SARS-CoV-1, monoclonal antibodies targeting SARS-CoV-1 RBD failed to bind to SARS-CoV-2-RBD [9]. However, in early research, sera from convalescent SARS patients were found to inhibit SARS-CoV-2 viral entry *in vitro*, albeit with lower efficiency than it inhibited SARS-CoV-1 [28].

Comparative genomic analysis reveals that several regions of the coronavirus genome are likely critical to virulence. The S1 domain of the spike protein, which contains the receptor binding motif, evolves more rapidly than the S2 domain [18,19]. However, even within the S1 domain, some regions are more conserved than others, with the receptors in S1's N-terminal domain (S1-NTD) evolving more rapidly than those in its C-terminal domain (S1-CTD) [19]. Both S1-NTD and S1-CTD are involved in receptor binding and can function as RBDs to bind proteins and sugars [18], but RBDs in the S1-NTD typically bind to sugars, while those in the S1-CTD recognize protein receptors [5]. Viral receptors show higher affinity with protein receptors than sugar receptors [5], which suggests that positive selection on or relaxed conservation of the S1-NTD might reduce the risk of a deleterious mutation that would prevent binding. The SARS-CoV-2 S protein also contains an RRAR furin recognition site at the S1/S2 junction [9,20], setting it apart from both bat coronavirus RaTG13, with which it shares 96% genome sequence identity, and SARS-CoV-1 [202]. Such furin cleavage sites are commonly found in highly virulent influenza viruses [203,204]. The furin recognition site at the S1/S2 junction is likely to increase pathogenicity via destabilization of the spike protein during fusion to ACE2 and the facilitation of cell-cell adhesion [9,20,37,201,203,204]. These factors may influence the virulence of SARS-CoV-2 relative to other beta coronaviruses. Additionally, a major concern has been the emergence of SARS-CoV-2 variants with increased virulence. The extent to which evolution within SARS-CoV-2 may affect pathogenesis is reviewed below.

# 0.8 Molecular Signatures, Transmission, and Variants of Concern

Genetic variation in SARS-CoV-2 has been used to elucidate patterns over time and space. Many mutations are neutral in their effect and can be used to trace transmission patterns. Such signatures within SARS-CoV-2 have provided insights during outbreak investigations [205,206,207]. Similar mutations observed in several patients may indicate that the patients belong to the same transmission group. The tracking of SARS-CoV-2 mutations is recognized as an essential tool for controlling future outbreaks and tracing the path of the spread of SARS-CoV-2. In the first months of the pandemic in early 2020, early genomic surveillance efforts in Guangdong, China revealed that local transmission rates were low and that most cases arising in the province were imported [208]. Since then, efforts have varied widely among countries: for example, the U.K. has coordinated a national database of viral genomes [209], but efforts to collect this type of data in the United States have been more limited [210]. Studies have applied phylogenetic analyses of viral genomes to determine the source of local COVID-19 outbreaks in Connecticut (USA), [211], the New York City area (USA) [212], and Iceland [213]. There has been an ongoing effort to collect SARS-CoV-2 genomes throughout the COVID-19 outbreak, and as of summer 2021, millions of genome sequences have been collected from patients. The sequencing data can be found at GISAID [214], NCBI [215], and COVID-19 data portal [216].

Ongoing evolution can be observed in genomic data collected through molecular surveillance efforts. In some cases, mutations can produce functional changes that can impact pathogenesis. One early example is the spike protein mutation D614G, which appeared in March 2020 and became dominant worldwide by the end of May 2020 [217,218]. This variant was associated with increased infectivity and increased viral load, but not with more severe disease outcomes [217,219]. This increased virulence is likely achieved by altering the conformation of the S1 domain to facilitate binding to ACE2 [219]. Similarly, the N439K mutation within the RBD of the spike protein is likely associated with increased transmissibility and enhanced binding affinity for hACE2, although it is also not thought to affect disease outcomes [220]. In contrast, a mutation in ORF8 that was identified in Singapore in the early months of 2020 was associated with cases of COVID-19 that were less likely to require treatment with supplemental oxygen [221], and a deletion surrounding the furin site insertion at the S1/S2 boundary has been identified only rarely in clinical settings [222], suggesting that these mutations may disadvantage viral pathogenesis in human hosts. Thus, mutations have been associated with both virological and clinical differences in pathogenesis.

Several VOCs have also been identified and designated through molecular surveillance efforts [223]. The Alpha variant (lineage B.1.1.7) was first observed in the U.K. in October 2020 before it quickly spread around the world [224]. Other variants meriting further investigation have also been identified, including the Beta variant (B.1.351 lineage) first identified in South Africa and the Gamma variant (P.1 lineage) initially associated with outbreaks in Brazil. These lineages share independently acquired mutations that may affect pathogenicity [225,226,227,228,229]. For example, they are all associated with a greater binding affinity for hACE2 than that of the wildtype variant [227,230,231], but they were not found to have more efficient cell entry than the wildtype virus [232]. A fourth VOC, the Delta variant (B.1.617.2 and AY.1, AY.2, and AY.3 lineages), was identified in India in late 2020 [233]. Some of the mutations associated with this lineage may alter fusogenicity and enhance furin cleavage, among other effects associated with increased pathogenicity [234]. The changes in these VOC demonstrate how ongoing evolution in SARS-CoV-2 can drive changes in how the virus interacts with host cells.

## 0.9 Quantifying Viral Presence

Assessing whether a virus is present in a sample is a more complex task than it initially seems. Many diagnostic tests rely on real-time polymerase chain reaction (RT-PCR) to test for the presence versus absence of a virus [235]. They may report the cycle threshold ($C_t$) indicating the number of doubling cycles required for the target (in this case, SARS-CoV-2) to become detectable. A lower $C_t$ therefore corresponds to a higher viral load. The $C_t$ that corresponds to a positive can vary widely, but is often around 35. This information is sufficient to answer many questions, since an amplicon must be present in order to be duplicated in RT-PCR. For example, if a patient is presenting with COVID-19 symptoms, a positive RT-PCR test can confirm the diagnosis.

However, RT-PCR analysis alone cannot provide the information needed to determine whether a virus is present at sufficient levels to be infectious [236]. Some studies have therefore taken the additional step of cultivating samples *in vitro* in order to observe whether cells become infected with SARS-CoV-2. One study collected upper respiratory tract samples from COVID-19 patients, analyzed them with RT-PCR to determine the cycle threshold, and then attempted to cultivate the SARS-CoV-2 virus in VeroE6 cells [236]. This study found that out of 246 samples, less than half (103) produced a positive culture. Moreover, at a $C_t$ of 35, only 5 out of 60 samples grew *in vitro*. Therefore, the RT-PCR-confirmed presence of SARS-CoV-2 in a sample does not necessarily indicate that the virus is present at a high enough concentration to grow and/or spread.

## 0.10 Mechanisms of Transmission

When a human host is infected with a virus and is contagious, person-to-person viral transmission can occur through several possible mechanisms. When a contagious individual sneezes, coughs, or exhales, they produce respiratory droplets that can contain a large number of viral particles [237]. Viral particles can enter the body of a new host when they then come in contact with the oral, nasal, eye, or other mucus membranes [237]. The primary terms typically used to discuss the transmission of viruses via respiratory droplets are droplet, aerosol, and contact transmission [238]. The distinction between droplet and aerosol transmission is typically anchored on whether a particle containing the virus is larger or smaller than 5 micrometers (μm) [239,240]. Droplet transmission typically refers to contact with large droplets that fall quickly to the ground at close range, such as breathing in droplets produced by a sneeze [237,239]. Aerosol transmission typically refers to much smaller particles (less than 5 μm) produced by sneezing, coughing, or exhaling [237,238] that can remain suspended over a longer period of time and potentially be moved by air currents [237]. It is also possible that viral particles deposited on surfaces via large respiratory droplets could later be aerosolized [237]. The transmission of viral particles that have settled on a surface is typically referred to as contact or fomite transmission [237,241]. Any respiratory droplets that settle on a surface could contribute to fomite transmission [237]. Droplet and contact transmission are both well-accepted modes of transmission for many viruses associated with common human illnesses, including influenza and rhinovirus [237].

The extent to which aerosol transmission contributes to the spread of respiratory viruses is more widely debated. In influenza A, for example, viral particles can be detected in aerosols produced by infected individuals, but it is not clear to what extent these particles drive the spread of influenza A infection [237,238,242,243,244]. Regardless of its role in the spread of influenza A, however, aerosol transmission likely played a role in outbreaks such as the 1918 Spanish Influenza (H1N1) and 2009 "swine flu" (pH1N1) [244]. All three of these mechanisms have been identified as possible contributors to the transmission of HCoVs [237], including the highly pathogenic coronaviruses SARS-CoV-1 and MERS-CoV [245,246]. Transmission of SARS-CoV-1 is thought to proceed primarily through droplet transmission, but aerosol transmission is also considered possible [237,247,248], and fomite transmission may have also played an important role in some outbreaks [249]. Similarly, the primary mechanism of MERS transmission is thought to be droplets because inter-individual transmission appears to be associated with close interpersonal contact (e.g., household or healthcare settings), but aerosolized particles of the MERS virus have been reported to persist much more robustly than influenza A under a range of environmental conditions [250,251]. However, few of these analyses have sought to grow positive samples in culture and thus to confirm their potential to infect new hosts.

Contact, droplet, and aerosol transmission are therefore all worth evaluating when considering possible modes of transmission for a respiratory virus like SARS-CoV-2. The stability of the SARS-CoV-2 virus both in aerosols and on a variety of surfaces was found to be similar to that of SARS-CoV-1 [252]. Droplet-based and contact transmission were initially put forward as the greatest concern for the spread of SARS-CoV-2 [253], with droplet transmission considered the dominant mechanism driving the spread of the virus [254] because the risk of fomite transmission under real-world conditions is likely to be substantially lower than the conditions used for experimental analyses [255]. The COVID-19 pandemic has, however, exposed significant discrepancies in how terms pertaining to airborne viral particles are interpreted in different contexts [239]. The 5-µm distinction between "droplets" and "aerosols" is typical in the biological literature but is likely an artifact of historical science rather than a meaningful boundary in biology or physics [240]. Additionally, various ambient conditions such as air flow can influence how particles of different sizes fall or spread [239]. Despite initial skepticism about airborne transmission of SARS-CoV-2 through small particles [240], evidence now suggests that small particles can contribute to SARS-CoV-2 transmission [252,256,257,258]. For example, one early study detected SARS-CoV-2 viral particles in air samples taken from hospitals treating COVID-19 patients, although the infectivity of these samples was not assessed [259]. Subsequently, other studies have been successful in growing SARS-CoV-2 in culture with samples taken from the air [260,261] while others have not [262,263] (see [264] for a systematic review of available findings as of July 2020). The fact that viable SARS-CoV-2 may exist in aerosolized particles calls into question whether some axioms of COVID-19 prevention, such as 2-meter social distancing, are sufficient [240,260,265].

## 0.10.1 Symptoms and Viral Spread

Other aspects of pathogenesis are also important to understanding how the virus spreads, especially the relationship between symptoms, viral shedding, and contagiousness. Symptoms associated with reported cases of COVID-19 range from mild to severe [1], but some individuals who contract COVID-19 remain asymptomatic throughout the duration of the illness [266]. The incubation period, or the time period between exposure and the onset of symptoms, has been estimated at five to eight days, with means of 4.91 (95% confidence interval (CI) 4.35-5.69) and 7.54 (95% CI 6.76-8.56) reported in two different Asian cities and a median of 5 (IQR 1 to 6) reported in a small number of patients in a Beijing hospital [267,268].

However, the exact relationship between contagiousness and viral shedding remains unclear. Estimates suggest that viral shedding can, in some cases, begin as early as 12.3 days (95% CI 5.9-17.0) before the onset of symptoms, although this was found to be very rare, with less than 0.1% of transmission events occurring 7 or more days before symptom onset [269]. Transmissibility appeared to peak around the onset of symptoms (95% CI -0.9 - 0.9 days), and only 44% (95% CI 30–57%) of transmission events were estimated to occur from presymptomatic contacts [269]. A peak in viral load corresponding to the onset of symptoms was also confirmed by another study [236]. As these trends became apparent, concerns arose due to the potential for individuals who did not yet show symptoms to transmit the virus [270]. Recovered individuals may also be able to transmit the virus after their symptoms cease. A study of the communicable period based on twenty-four individuals who tested positive for SARS-CoV-2 prior to or without developing symptoms estimated that individuals may be contagious for one to twenty-one days, but they note that this estimate may be low [266]. In an early study, viral nucleic acids were reported to remain at observable levels in the respiratory specimens of recovering hospitalized COVID-19 patients for a median of 20 days and with a maximum observed duration through 37 days, when data collection for the study ceased [78].

As more estimates of the duration of viral shedding were released, they converged around approximately three weeks from first positive PCR test and/or onset of symptoms (which, if present, are usually identified within three days of the initial PCR test). For example, in some studies, viral shedding was reported for up to 28 days following symptom onset [271] and for one to 24 days from first positive PCR test, with a median of 12 days [67]. On the other hand, almost 70% of patients were reported to still have symptoms at the time that viral shedding ceased, although all symptoms reduced in prevalence between onset and cessation of viral shedding [272]. The median time that elapsed between the onset of symptoms and cessation of viral RNA shedding was 23 days and between first positive PCR test and cessation of viral shedding was 17 days [272]. The fact that this study reported symptom onset to predate the first positive PCR test by an average of three days, however, suggests that there may be some methodological differences between it and related studies. Furthermore, an analysis of residents of a nursing home with a known SARS-CoV-2 case measured similar viral load in residents who were asymptomatic regardless of whether they later developed symptoms, and the load in the asymptomatic residents was comparable to that of residents who displayed either typical or atypical symptoms [273]. Taken together, these results suggest that the presence or absence of symptoms are not reliable predictors of viral shedding or of SARS-CoV-2 status (e.g, [274]). However, it should be noted that viral shedding is not necessarily a robust indicator of contagiousness. The risk of spreading the infection was low after ten days from the onset of symptoms, as viral load in sputum was found to be unlikely to pose a significant risk based on efforts to culture samples *in vitro* [271]. The relationship between symptoms, detectable levels of the virus, and risk of viral spread is therefore complex.

The extent to which asymptomatic or presymptomatic individuals are able to transmit SARS-CoV-2 has been a question of high scientific and community interest. Early reports (February and March 2020) described transmission from presymptomatic SARS-CoV-2-positive individuals to close family contacts [275,276]. One of these reports [276] also included a description of an individual who tested positive for SARS-CoV-2 but never developed symptoms. Later analyses also sought to estimate the proportion of infections that could be traced back to a presymptomatic or asymptomatic individual (e.g., [277]). Estimates of the proportion of individuals with asymptomatic infections have varied widely. The proportion of asymptomatic individuals on board the Diamond Princess cruise ship, which was the site of an early COVID-19 outbreak, was estimated at 17.9% [278]. In contrast, a model using the prevalence of antibodies among residents of Wuhan, China estimated a much higher rate of asymptomatic cases, at approximately 7 in 8, or 87.5% [279]. An analysis of the populations of care homes in London found that, among the residents (median age 85), the rate of asymptomatic infection was 43.8%, and among the caretakers (median age 47), the rate was 49.1% [280]. The duration of viral shedding may also be longer in individuals with asymptomatic cases of COVID-19 compared to those who do show symptoms [281]. As a result, the potential for individuals who do not know they have COVID-19 to spread the virus raises significant concerns. In Singapore and Tianjin, two cities studied to estimate incubation period, an estimated 40-50% and 60-80% of cases, respectively, were considered to be caused by contact with asymptomatic individuals [267]. An analysis of viral spread in the Italian town of Vo', which was the site of an early COVID-19 outbreak, revealed that 42.5% of cases were asymptomatic and that the rate was similar across age groups [282]. The argument was thus made that the town's lockdown was imperative for controlling the spread of COVID-19 because it isolated asymptomatic individuals. While more models are likely to emerge to better explore the effect of asymptomatic individuals on SARS-CoV-2 transmission, these results suggest that strategies for identifying and containing asymptomatic but contagious individuals are important for managing community spread.

## 0.10.2    Estimating the Fatality Rate

Estimating the occurrence of asymptomatic and mild COVID-19 cases is important to identifying the mortality rate associated with COVID-19. The mortality rate of greatest interest would be the total number of fatalities as a fraction of the total number of people infected. One commonly reported metric is the case fatality rate (CFR), which compares the number of COVID-19 related deaths to the number of confirmed or suspected cases. However, in locations without universal testing protocols, it is impossible to identify all infected individuals because so many asymptomatic or mild cases go undetected. Therefore, a more informative metric is the infection fatality rate (IFR), which compares the known deaths to the estimated number of cases. It thus requires the same numerator as CFR, but divides by an approximation of the total number of cases rather than only the observed/suspected cases. IFR varies regionally, with some locations observed to have IFRs as low as 0.17% while others are as high as 1.7% [283].

Estimates of CFR at the national and continental level and IFR at the continent level is maintained by the Centre for Evidence-Based Medicine [284]. Several meta-analyses have also sought to estimate IFR at the global scale. These estimates have varied; one peer-reviewed study aggregated data from 24 other studies and estimated IFR at 0.68% (95% CI 0.53%–0.82%), but a preprint that aggregated data from 139 countries calculated a global IFR of 1.04% (95% CI 0.77%-1.38%) when false negatives were considered in the model [283,285]. A similar prevalence estimate was identified through a repeated cross-sectional serosurvey conducted in New York City that estimated the IFR as 0.97% [286]. Examination of serosurvey-based estimates of IFR identified convergence on a global IFR estimate of 0.60% (95% CI 0.42%–0.77%) [283]. All of these studies note that IFR varies widely by location, and it is also expected to vary with demographic and health-related variables such as age, sex, prevalence of comorbidities, and access to healthcare and testing [287]. Estimates of infection rates are becoming more feasible as more data becomes available for modeling and will be bolstered as serological testing becomes more common and more widely available. However, this research may be complicated due to the emergence of variants over time, as well as the varying availability and acceptance of vaccines in different communities and locations.

# 0.11 Dynamics of Transmission

Disease spread dynamics can be estimated using $R_0$, the basic reproduction number, and $R_t$, the effective reproduction number. Accurate estimates of both are crucial to understanding the dynamics of infection and to predicting the effects of different interventions. $R_0$ is the average number of new (secondary) infections caused by one infected person, assuming a wholly susceptible population [288], and is one of the most important epidemiological parameters [289]. A simple mechanistic model used to describe infectious disease dynamics is a susceptible-infected-recovered compartmental model [290,291]. In this model, individuals move through three states: susceptible, infected, and recovered; two parameters, $\gamma$ and $\beta$, specify the rate at which the infectious recover, and the infection transmission rate, respectively, and $R_0$ is estimated as the ratio of $\beta$ and $\gamma$ [289,292]. A pathogen can invade a susceptible population only if $R_0 > 1$ [289,293]. The spread of an infectious disease at a particular time t can be quantified by $R_t$, the effective reproduction number, which assumes that part of the population has already recovered (and thus gained immunity to reinfection) or that mitigating interventions have been put into place. For example, if only a fraction $S_t$ of the population is still susceptible, $R_t = S_t \times R_0$. When $R_t$ is greater than 1, an epidemic grows (i.e., the proportion of the population that is infectious increases); when $R_t$ is less than 1, the proportion of the population that is infectious decreases. $R_0$ and $R_t$ can be estimated directly from epidemiological data or inferred using susceptible-infected-recovered-type models. To capture the dynamics of SARS-CoV-2 accurately, the addition of a fourth compartment, i.e. a susceptible-exposed-infectious-recovered model, may be appropriate because such models account for the relative lengths of incubation and infectious periods [294].

Original estimates of $R_0$ for COVID-19 lie in the range $R_0$=1.4-6.5 [295,296,297]. Variation in $R_0$ is expected between different populations, and the estimated values of $R_0$ discussed below are for specific populations in specific environments. The different estimates of $R_0$ should not necessarily be interpreted as a range of estimates of the same underlying parameter. In one study of international cases, the predicted value was $R_0$=1.7 [298]. In China (both Hubei province and nationwide), the value was predicted to lie in the range $R_0$=2.0-3.6 [295,299,300]. Another estimate based on a cruise ship where an outbreak occurred predicted $R_0$=2.28 [301]. Susceptible-exposed-infectious-recovered model-derived estimates of $R_0$ range from 2.0 - 6.5 in China [302,303,304,305] to $R_0$=4.8 in France [306]. Using the same model as for the French population, a study estimated $R_0$=2.6 in South Korea [306], which is consistent with other studies [307]. From a meta-analysis of studies estimating $R_0$, [296] the median $R_0$ was estimated to be 2.79 (IQR 1.16) based on twelve studies published between January 1 and February 7, 2020.

Inference of the effective reproduction number can provide insight into how populations respond to an infection and the effectiveness of interventions. In China, $R_t$ was predicted to lie in the range 1.6-2.6 in January 2020, before travel restrictions [308]. $R_t$ decreased from 2.35 one week before travel restrictions were imposed (January 23, 2020), to 1.05 one week after. Using their model, the authors also estimated the probability of new outbreaks occurring. Assuming individual-level variation in transmission comparable to that of MERS or SARS, the probability of a single individual exporting the virus and causing a large outbreak is 17-25%, and assuming variation like that of SARS and transmission patterns like those observed for COVID-19 in Wuhan, the probability of a large outbreak occurring after ≥4 infections exist at a new location is greater than 50%. An independent study came to similar conclusions, finding $R_t$=2.38 in the two-week period before January 23 with a decrease to $R_t$ = 1.34 (using data from January 24 to February 3) or $R_t$=0.98 (using data from January 24 to February 8) [297]. In South Korea, $R_t$ was inferred for February through March 2020 in two cities, Daegu (the center of the outbreak) and Seoul [307]. Metro data was also analyzed to estimate the effects of social distancing measures. $R_t$ decreased in Daegu from around 3 to <1 over the period that social distancing measures were introduced. In Seoul, $R_t$ decreased slightly, but remained close to 1 (and larger than $R_t$ in Daegu). These findings indicate that social distancing measures appeared to be effective in containing the infection in Daegu, but in Seoul, $R_t$ remained above 1, meaning secondary outbreaks remained possible. The study also shows the importance of region-specific analysis: the large decline in case load nationwide was mainly due to the Daegu region and could mask persistence of the epidemic in other regions, such as Seoul and Gyeonggi-do. In Iran, estimates of $R_t$ declined from 4.86 in the first week to 2.1 by the fourth week after the first cases were reported [309]. In Europe, analysis of 11 countries inferred the dynamics of $R_t$ over a time range from the beginning of the outbreak until March 28, 2020, by which point most countries had implemented major interventions (such as stay-at-home orders, public gathering bans, and school closures) [310]. Across all countries, the mean $R_t$ before interventions began was estimated as 3.87; $R_t$ varied considerably, from below 3 in Norway to above 4.5 in Spain. After interventions, $R_t$ decreased by an average of 64% across all countries, with mean $R_t$=1.43. The lowest predicted value was 0.97 for Norway and the highest was 2.64 for Sweden, which could be related to the fact that Sweden did not implement social distancing measures on the same scale as other countries. The study concludes that while large changes in $R_t$ are observed, it is too early to tell whether the interventions put into place are sufficient to decrease $R_t$ below 1.

Evolution within SARS-CoV-2 has also driven changes in the estimated reproduction number for different populations at different times. As of June 2021, the reproduction number had increased globally relative to 2020, and increased transmissibility over the wildtype variant was observed for the Alpha, Beta, Gamma, and Delta VOC [311]. In the U.S. between December 2020 and January 2021, B.1.1.7 (Alpha) was estimated to have an increased transmission of 35 to 45% relative to common SARS-CoV-2 variants at the time, with B.1.1.7 the dominant SARS-CoV-2 variant in some places at some timepoints [312]. This lineage was estimated to have increased transmissibility of 43 to 90% in the U.K. [313]. An estimate of the reproduction number of B.1.1.7 in the U.K. from September to December 2020 yielded 1.59 overall and between 1.56 and 1.95 in different regions of the country [229]. The Delta variant is particularly transmissible, and it has been estimated to be twice as transmissible than the wildtype variant of SARS-CoV-2 [311]. A review of the literature describing the Delta variant identified a mean estimated $R_0$ of 5.08 [314]. Such differences can affect fitness and therefore influence the relative contributions of different lineages to a given viral gene pool over time [315]. Therefore, the evolution of the virus can result in shifts in the reproduction rate.

More generally, population-level epidemic dynamics can be both observed and modeled [292]. Data and empirically determined biological mechanisms inform models, while models can be used to try to understand data and systems of interest or to make predictions about possible future dynamics, such as the estimation of capacity needs [316] or the comparison of predicted outcomes among prevention and control strategies [317,318]. Many current efforts to model $R_t$ have also led to tools that assist the visualization of estimates in real time or over recent intervals [319,320]. These are valuable resources, yet it is also important to note that the estimates arise from models containing many assumptions and are dependent on the quality of the data they use, which varies widely by region.

# 0.12 Conclusions

The novel coronavirus SARS-CoV-2 is the third HCoV to emerge in the 21st century, and research into previous HCoVs has provided a strong foundation for characterizing the pathogenesis and transmission of SARS-CoV-2. Critical insights into how the virus interacts with human cells have been gained from previous research into HCoVs and other viral infections. With the emergence of three devastating HCoV over the past twenty years, emergent viruses are likely to represent an ongoing threat. Contextualizing SARS-CoV-2 alongside other viruses serves not only to provide insights that can be immediately useful for combating this virus itself but may also prove valuable in the face of future viral threats.

Host-pathogen interactions provide a basis not only for understanding COVID-19, but also for developing a response. As with other HCoVs, the immune response to SARS-CoV-2 is likely driven by detection of its spike protein, which allows it to enter cells through ACE2. Epithelial cells have also emerged as the major cellular target of the virus, contextualizing the respiratory and gastrointestinal symptoms that are frequently observed in COVID-19. Many of the mechanisms that facilitate the pathogenesis of SARS-CoV-2 are currently under consideration as possible targets for the treatment or prevention of COVID-19 [10,11]. Research in other viruses also provides a foundation for understanding the transmission of SARS-CoV-2 among people and can therefore inform efforts to control the virus's spread. Airborne forms of transmission (droplet and aerosol transmission) have emerged as the primary modes by which the virus spreads to new hosts. Asymptomatic transmission was also a concern in the SARS outbreak of 2002-03 and, as in the current pandemic, presented challenges for estimating rates of infection [321]. These insights are important for developing a public health response, such as the CDC's shift in its recommendations surrounding masking [322].

Even with the background obtained from research in SARS and MERS, COVID-19 has revealed itself to be a complex and difficult-to-characterize disease that has many possible presentations that vary with age. Variability in presentation, including cases with no respiratory symptoms or with no symptoms altogether, were also reported during the SARS epidemic at the beginning of the 21st century [321]. The variability of both which symptoms present and their severity have presented challenges for public health agencies seeking to provide clear recommendations regarding which symptoms indicate SARS-CoV-2 infection and should prompt isolation. Asymptomatic cases add complexity both to efforts to estimate statistics such as $R_0$ and $R_t$, which are critical to understanding the transmission of the virus, and IFR, which is an important component of understanding its impact on a given population. The development of diagnostic technologies over the course of the pandemic has facilitated more accurate identification, including of asymptomatic cases [235]. As more cases have been diagnosed, the health conditions and patient characteristics associated with more severe infection have also become more clear, although there are likely to be significant sociocultural elements that also influence these outcomes [323]. While many efforts have focused on adults, and especially older adults because of the susceptibility of this demographic, additional research is needed to understand the presentation of COVID-19 and MIS-C in pediatric patients. As more information is uncovered about the pathogenesis of HCoV and SARS-CoV-2 specifically, the diverse symptomatology of COVID-19 has and likely will continue to conform with the ever-broadening understanding of how SARS-CoV-2 functions within a human host.

While the SARS-CoV-2 virus is very similar to other HCoV in several ways, including in its genomic structure and the structure of the virus itself, there are also some differences that may account for differences in the COVID-19 pandemic compared to the SARS and MERS epidemics of the past two decades. The $R_0$ of SARS-CoV-2 has been estimated to be similar to SARS-CoV-1 but much higher than that of MERS-CoV [324], although a higher $R_0$ has been estimated for some VOC. While the structures of the viruses are very similar, evolution among these species may account for differences in their transmissibility and virulence. For example, the acquisition of a furin cleavage site the S1/S2 boundary within the SARS-CoV-2 S protein may be associated with increased virulence. Additionally, concerns have been raised about the accumulation of mutations within the SARS-CoV-2 species itself, and whether these could influence virulence [325]. These novel variants may be resistant to vaccines and antibody treatments such as Bamlanivimab that were designed based on the wildtype spike protein [10,326]. As a consequence of reliance on targeting the SARS-CoV-2 spike protein for many therapeutic and prophylactic strategies, increased surveillance is required to rapidly identify and prevent the spread of novel SARS-CoV-2 variants with alterations to the spike protein. The coming of age of genomic technologies has made these types of analyses feasible, and genomics research characterizing changes in SARS-CoV-2 along with temporal and spatial movement is likely to provide additional insights into whether within-species evolution influences the effect of the virus on the human host. Additionally, the rapid development of sequencing technologies over the past decade has made it possible to rapidly characterize the host response to the virus. For example, proteomics analysis of patient-derived cells revealed candidate genes whose regulation is altered by SARS-CoV-2 infection, suggesting possible approaches for pharmaceutical invention and providing insight into which systems are likely to be disrupted in COVID-19 [190]. As more patient data becomes available, the biotechnological advances of the 2000s are expected to allow for more rapid identification of potential drug targets than was feasible during the SARS, or even MERS, pandemic.

Thus, the COVID-19 crisis continues to evolve, but the insights acquired over the past 20 years of HCoV research have provided a solid foundation for understanding the SARS-CoV-2 virus and the disease it causes. As the scientific community continues to respond to COVID-19 and to elucidate more of the relationships between pathogenesis, transmission, host regulatory responses, and symptomatology, this understanding will no doubt continue to evolve and to reveal additional connections among virology, pathogenesis, and health. This review represents a collaboration between scientists from diverse backgrounds to contextualize this virus at the union of many different biological disciplines [327]. At present, understanding the SARS-CoV-2 virus and its pathogenesis is critical to a holistic understanding of the COVID-19 pandemic. In the future, interdisciplinary work on SARS-CoV-2 and COVID-19 may guide a response to a new viral threat.

# 1   Additional Items

## 1.1  Competing Interests

| Author | Competing Interests | Last Reviewed |
|---|---|---|
| Halie M. Rando | None | 2021-01-20 |
| Adam L. MacLean | None | 2021-02-23 |
| Alexandra J. Lee | None | 2020-11-09 |
| Ronan Lordan | None | 2020-11-03 |
| Sandipan Ray | None | 2020-11-11 |
| Vikas Bansal | None | 2021-01-25 |
| Ashwin N. Skelly | None | 2020-11-11 |
| Elizabeth Sell | None | 2020-11-11 |
| John J. Dziak | None | 2020-11-11 |
| Lamonica Shinholster | None | 2020-11-11 |
| Lucy D'Agostino McGowan | Received consulting fees from Acelity and Sanofi in the past five years | 2020-11-10 |
| Marouen Ben Guebila | None | 2021-08-02 |
| Nils Wellhausen | None | 2020-11-03 |
| Sergey Knyazev | None | 2020-11-11 |

| Author | Competing Interests | Last Reviewed |
|---|---|---|
| Simina M. Boca | Currently an employee at AstraZeneca, Gaithersburg, MD, USA, may own stock or stock options. | 2021-07-01 |
| Stephen Capone | None | 2020-11-11 |
| Yanjun Qi | None | 2020-07-09 |
| YoSon Park | YoSon Park is affiliated with Pfizer Worldwide Research. The author has no financial interests to declare and contributed as an author prior to joining Pfizer, and the work was not part of a Pfizer collaboration nor was it funded by Pfizer. | 2020-01-22 |
| Yuchen Sun | None | 2020-11-11 |
| David Mai | None | 2021-01-08 |
| Joel D Boerckel | None | 2021-03-26 |
| Christian Brueffer | Employee and shareholder of SAGA Diagnostics AB. | 2020-11-11 |
| James Brian Byrd | Funded by FastGrants to conduct a COVID-19-related clinical trial | 2020-11-12 |
| Jeremy P. Kamil | TBD | 2021-04-30 |
| Jinhui Wang | None | 2021-01-21 |
| Ryan Velazquez | None | 2020-11-10 |
| Gregory L Szeto | None | 2020-11-16 |
| John P. Barton | None | 2020-11-11 |
| Rishi Raj Goel | None | 2021-01-20 |
| Serghei Mangul | None | 2020-11-11 |
| Tiago Lubiana | None | 2020-11-11 |
| COVID-19 Review Consortium | None | 2021-01-16 |
| Anthony Gitter | Filed a patent application with the Wisconsin Alumni Research Foundation related to classifying activated T cells | 2020-11-10 |
| Casey S. Greene | None | 2021-01-20 |

## 1.2 Author Contributions

| Author | Contributions |
|---|---|
| Halie M. Rando | Project Administration, Writing - Original Draft, Writing - Review & Editing |
| Adam L. MacLean | Writing - Original Draft, Writing - Review & Editing |
| Alexandra J. Lee | Writing - Original Draft, Writing - Review & Editing |
| Ronan Lordan | Writing - Review & Editing |

| Author | Contributions |
| --- | --- |
| Sandipan Ray | Writing - Original Draft, Writing - Review & Editing |
| Vikas Bansal | Writing - Original Draft, Writing - Review & Editing |
| Ashwin N. Skelly | Writing - Original Draft, Writing - Review & Editing |
| Elizabeth Sell | Writing - Original Draft, Writing - Review & Editing |
| John J. Dziak | Writing - Original Draft, Writing - Review & Editing |
| Lamonica Shinholster | Writing - Original Draft |
| Lucy D'Agostino McGowan | Writing - Original Draft, Writing - Review & Editing |
| Marouen Ben Guebila | Writing - Original Draft, Writing - Review & Editing |
| Nils Wellhausen | Visualization, Writing - Original Draft, Writing - Review & Editing |
| Sergey Knyazev | Writing - Original Draft, Writing - Review & Editing |
| Simina M. Boca | Writing - Review & Editing |
| Stephen Capone | Writing - Original Draft, Writing - Review & Editing |
| Yanjun Qi | Visualization |
| YoSon Park | Writing - Original Draft, Writing - Review & Editing |
| Yuchen Sun | Visualization |
| David Mai | Writing - Original Draft, Writing - Review & Editing |
| Joel D Boerckel | Writing - Review & Editing |
| Christian Brueffer | Writing - Original Draft, Writing - Review & Editing |
| James Brian Byrd | Writing - Original Draft, Writing - Review & Editing |
| Jeremy P. Kamil | Writing - Review & Editing |
| Jinhui Wang | Writing - Review & Editing |
| Ryan Velazquez | Writing - Review & Editing |
| Gregory L Szeto | Writing - Review & Editing |
| John P. Barton | Writing - Original Draft, Writing - Review & Editing |
| Rishi Raj Goel | Writing - Original Draft, Writing - Review & Editing |
| Serghei Mangul | Writing - Review & Editing |
| Tiago Lubiana | Writing - Review & Editing |
| COVID-19 Review Consortium | Project Administration |
| Anthony Gitter | Methodology, Project Administration, Software, Writing - Review & Editing |
| Casey S. Greene | Conceptualization, Software, Writing - Review & Editing |

# 1.3  Acknowledgements

We thank Nick DeVito for assistance with the Evidence-Based Medicine Data Lab COVID-19 TrialsTracker data. We thank Yael Evelyn Marshall who contributed writing (original draft) as well as reviewing and editing of pieces of the text but who did not formally approve the manuscript, as well as Ronnie Russell, who contributed text to and helped develop the structure of the manuscript early in the writing process and Matthias Fax who helped with writing and editing text related to diagnostics. We are also very grateful to James Fraser for suggestions about successes and limitations in the area of computational screening for drug repurposing. We are grateful to the following contributors for reviewing pieces of the text: Nadia Danilova, James Eberwine and Ipsita Krishnan.

*Mortality Weekly Report* (2020-07-17) https://doi.org/gg8r2m DOI: 10.15585/mmwr.mm6928a2 · PMID: 32673296 · PMCID: PMC7366851

82. **Population-scale longitudinal mapping of COVID-19 symptoms, behaviour and testing** William E Allen, Han Altae-Tran, James Briggs, Xin Jin, Glen McGee, Andy Shi, Rumya Raghavan, Mireille Kamariza, Nicole Nova, Albert Pereta, … Xihong Lin *Nature Human Behaviour* (2020-08-26) https://doi.org/gg9dfq DOI: 10.1038/s41562-020-00944-2 · PMID: 32848231 · PMCID: PMC7501153

83. **Extrapulmonary manifestations of COVID-19** Aakriti Gupta, Mahesh V Madhavan, Kartik Sehgal, Nandini Nair, Shiwani Mahajan, Tejasav S Sehrawat, Behnood Bikdeli, Neha Ahluwalia, John C Ausiello, Elaine Y Wan, … Donald W Landry *Nature Medicine* (2020-07-10) https://doi.org/gg4r37 DOI: 10.1038/s41591-020-0968-3 · PMID: 32651579

84. **Acute kidney injury in patients hospitalized with COVID-19** Jamie S Hirsch, Jia H Ng, Daniel W Ross, Purva Sharma, Hitesh H Shah, Richard L Barnett, Azzour D Hazzan, Steven Fishbane, Kenar D Jhaveri, Mersema Abate, … Jia Hwei Ng *Kidney International* (2020-07) https://doi.org/ggx24k DOI: 10.1016/j.kint.2020.05.006 · PMID: 32416116 · PMCID: PMC7229463

85. **COVAN is the new HIVAN: the re-emergence of collapsing glomerulopathy with COVID-19** Juan Carlos Q Velez, Tiffany Caza, Christopher P Larsen *Nature Reviews Nephrology* (2020-08-04) https://doi.org/gmmfx7 DOI: 10.1038/s41581-020-0332-3 · PMID: 32753739 · PMCID: PMC7400750

86. **Nervous system involvement after infection with COVID-19 and other coronaviruses** Yeshun Wu, Xiaolin Xu, Zijun Chen, Jiahao Duan, Kenji Hashimoto, Ling Yang, Cunming Liu, Chun Yang *Brain, Behavior, and Immunity* (2020-07) https://doi.org/ggq7s2 DOI: 10.1016/j.bbi.2020.03.031 · PMID: 32240762 · PMCID: PMC7146689

87. **Neurological associations of COVID-19** Mark A Ellul, Laura Benjamin, Bhagteshwar Singh, Suzannah Lant, Benedict Daniel Michael, Ava Easton, Rachel Kneen, Sylviane Defres, Jim Sejvar, Tom Solomon *The Lancet Neurology* (2020-09) https://doi.org/d259 DOI: 10.1016/s1474-4422(20)30221-0 · PMID: 32622375 · PMCID: PMC7332267

88. **How COVID-19 Affects the Brain** Maura Boldrini, Peter D Canoll, Robyn S Klein *JAMA Psychiatry* (2021-06-01) https://doi.org/gjj3cd DOI: 10.1001/jamapsychiatry.2021.0500 · PMID: 33769431

89. **Update on the neurology of COVID-19** Josef Finsterer, Claudia Stollberger *Journal of Medical Virology* (2020-06-02) https://doi.org/gg2qnn DOI: 10.1002/jmv.26000 · PMID: 32401352 · PMCID: PMC7272942

90. **COVID-19: A Global Threat to the Nervous System** Igor J Koralnik, Kenneth L Tyler *Annals of Neurology* (2020-06-23) https://doi.org/gg3hzh DOI: 10.1002/ana.25807 · PMID: 32506549 · PMCID: PMC7300753

Stratta, Vincenzo Cantaluppi *Critical Care* (2020-06-19) https://doi.org/gg35w7 DOI: 10.1186/s13054-020-03060-9 · PMID: 32560665 · PMCID: PMC7303575

100. **COVID-19 update: Covid-19-associated coagulopathy** Richard C Becker *Journal of Thrombosis and Thrombolysis* (2020-05-15) https://doi.org/ggwpp5 DOI: 10.1007/s11239-020-02134-3 · PMID: 32415579 · PMCID: PMC7225095

101. **The complement system in COVID-19: friend and foe?** Anuja Java, Anthony J Apicelli, MKathryn Liszewski, Ariella Coler-Reilly, John P Atkinson, Alfred HJ Kim, Hrishikesh S Kulkarni *JCI Insight* (2020-08-06) https://doi.org/gg4b5b DOI: 10.1172/jci.insight.140711 · PMID: 32554923 · PMCID: PMC7455060

102. **COVID-19, microangiopathy, hemostatic activation, and complement** Wen-Chao Song, Garret A FitzGerald *Journal of Clinical Investigation* (2020-06-22) https://doi.org/gg4b5c DOI: 10.1172/jci140183 · PMID: 32459663 · PMCID: PMC7410042

103. **Post-acute COVID-19 syndrome** Ani Nalbandian, Kartik Sehgal, Aakriti Gupta, Mahesh V Madhavan, Claire McGroder, Jacob S Stevens, Joshua R Cook, Anna S Nordvig, Daniel Shalev, Tejasav S Sehrawat, … Elaine Y Wan *Nature Medicine* (2021-03-22) https://doi.org/gjh7b4 DOI: 10.1038/s41591-021-01283-z · PMID: 33753937

104. **Six-Month Outcomes in Patients Hospitalized with Severe COVID-19** Leora I Horwitz, Kira Garry, Alexander M Prete, Sneha Sharma, Felicia Mendoza, Tamara Kahan, Hannah Karpel, Emily Duan, Katherine A Hochman, Himali Weerahandi *Journal of General Internal Medicine* (2021-08-05) https://doi.org/gmg6wv DOI: 10.1007/s11606-021-07032-9 · PMID: 34355349 · PMCID: PMC8341831

105. **Preliminary evidence on long COVID in children** Danilo Buonsenso, Daniel Munblit, Cristina De Rose, Dario Sinatti, Antonia Ricchiuto, Angelo Carfi, Piero Valentini *Acta Paediatrica* (2021-04-18) https://doi.org/gj9qd5 DOI: 10.1111/apa.15870 · PMID: 33835507 · PMCID: PMC8251440

106. **Characterizing long COVID in an international cohort: 7 months of symptoms and their impact** Hannah E Davis, Gina S Assaf, Lisa McCorkell, Hannah Wei, Ryan J Low, Yochai Re'em, Signe Redfield, Jared P Austin, Athena Akrami *EClinicalMedicine* (2021-08) https://doi.org/gmbdm7 DOI: 10.1016/j.eclinm.2021.101019 · PMID: 34308300 · PMCID: PMC8280690

107. **6-month consequences of COVID-19 in patients discharged from hospital: a cohort study** Chaolin Huang, Lixue Huang, Yeming Wang, Xia Li, Lili Ren, Xiaoying Gu, Liang Kang, Li Guo, Min Liu, Xing Zhou, … Bin Cao *The Lancet* (2021-01) https://doi.org/ghstsk DOI: 10.1016/s0140-6736(20)32656-8 · PMID: 33428867 · PMCID: PMC7833295

108. **Challenges in defining Long COVID: Striking differences across literature, Electronic Health Records, and patient-reported information** Halie M Rando, Tellen D Bennett, James Brian Byrd, Carolyn Bramante, Tiffany J Callahan, Christopher G Chute, Hannah E Davis, Rachel Deer, Joel Gagnier, Farrukh M Koraishy, … Melissa A Haendel *Cold*

… Petter Brodin *Cell* (2020-11) https://doi.org/d8fh DOI: 10.1016/j.cell.2020.09.016 · PMID: 32966765 · PMCID: PMC7474869

127. **Acute Heart Failure in Multisystem Inflammatory Syndrome in Children in the Context of Global SARS-CoV-2 Pandemic** Zahra Belhadjer, Mathilde Méot, Fanny Bajolle, Diala Khraiche, Antoine Legendre, Samya Abakka, Johanne Auriau, Marion Grimaud, Mehdi Oualha, Maurice Beghetti, … Damien Bonnet *Circulation* (2020-08-04) https://doi.org/ggwkv6 DOI: 10.1161/circulationaha.120.048360 · PMID: 32418446

128. **An adult with Kawasaki-like multisystem inflammatory syndrome associated with COVID-19** Sheila Shaigany, Marlis Gnirke, Allison Guttmann, Hong Chong, Shane Meehan, Vanessa Raabe, Eddie Louie, Bruce Solitar, Alisa Femia *The Lancet* (2020-07) https://doi.org/gg4sd6 DOI: 10.1016/s0140-6736(20)31526-9 · PMID: 32659211 · PMCID: PMC7351414

129. **Multisystem inflammatory syndrome in an adult following the SARS-CoV-2 vaccine (MIS-V)** Arvind Nune, Karthikeyan P Iyengar, Christopher Goddard, Ashar E Ahmed *BMJ Case Reports* (2021-07-29) https://doi.org/gmdmss DOI: 10.1136/bcr-2021-243888 · PMID: 34326117 · PMCID: PMC8323360

130. **COVID-19 associated Kawasaki-like multisystem inflammatory disease in an adult** Sabrina Sokolovsky, Parita Soni, Taryn Hoffman, Philip Kahn, Joshua Scheers-Masters *The American Journal of Emergency Medicine* (2021-01) https://doi.org/gg5tf4 DOI: 10.1016/j.ajem.2020.06.053 · PMID: 32631771 · PMCID: PMC7315983

131. **Case Report: Adult Post-COVID-19 Multisystem Inflammatory Syndrome and Thrombotic Microangiopathy** Idris Boudhabhay, Marion Rabant, Lubka T Roumenina, Louis-Marie Coupry, Victoria Poillerat, Armance Marchal, Véronique Frémeaux-Bacchi, Khalil El Karoui, Mehran Monchi, Franck Pourcine *Frontiers in Immunology* (2021-06-23) https://doi.org/gmdmst DOI: 10.3389/fimmu.2021.680567 · PMID: 34248962 · PMCID: PMC8260674

132. **Characteristics and Outcomes of US Children and Adolescents With Multisystem Inflammatory Syndrome in Children (MIS-C) Compared With Severe Acute COVID-19** Leora R Feldstein, Mark W Tenforde, Kevin G Friedman, Margaret Newhams, Erica Billig Rose, Heda Dapul, Vijaya L Soma, Aline B Maddux, Peter M Mourani, Cindy Bowens, … Overcoming COVID-19 Investigators *JAMA* (2021-03-16) https://doi.org/gh599q DOI: 10.1001/jama.2021.2091 · PMID: 33625505 · PMCID: PMC7905703

133. **Preliminary Evidence on Long COVID in children** Danilo Buonsenso, Daniel Munblit, Cristina De Rose, Dario Sinatti, Antonia Ricchiuto, Angelo Carfi, Piero Valentini *Cold Spring Harbor Laboratory* (2021-01-26) https://doi.org/fv9t DOI: 10.1101/2021.01.23.21250375

134. **Pediatric long-COVID: An overlooked phenomenon?** Caroline LH Brackel, Coen R Lap, Emilie P Buddingh, Marlies A Houten, Linda JTM Sande, Eveline J Langereis, Michiel AGE Bannier, Marielle WH Pijnenburg, Simone Hashimoto, Suzanne WJ Terheggen-Lagro *Pediatric*

*Infectious Diseases* (2021-08) https://doi.org/gsfg DOI: 10.1016/s1473-3099(21)00262-0 · PMID: 34022142 · PMCID: PMC8133765

228. **Evolution, correlation, structural impact and dynamics of emerging SARS-CoV-2 variants** Austin N Spratt, Saathvik R Kannan, Lucas T Woods, Gary A Weisman, Thomas P Quinn, Christian L Lorson, Anders Sönnerborg, Siddappa N Byrareddy, Kamal Singh *Computational and Structural Biotechnology Journal* (2021) https://doi.org/gk726x DOI: 10.1016/j.csbj.2021.06.037 · PMID: 34188776 · PMCID: PMC8225291

229. **Transmission of SARS-CoV-2 Lineage B.1.1.7 in England: Insights from linking epidemiological and genetic data** Erik Volz, Swapnil Mishra, Meera Chand, Jeffrey C Barrett, Robert Johnson, Lily Geidelberg, Wes R Hinsley, Daniel J Laydon, Gavin Dabrera, Áine O'Toole, … The COVID-19 Genomics UK (COG-UK) consortium *Cold Spring Harbor Laboratory* (2021-01-04) https://doi.org/ghrqv8 DOI: 10.1101/2020.12.30.20249034

230. **Experimental Evidence for Enhanced Receptor Binding by Rapidly Spreading SARS-CoV-2 Variants** Charlie Laffeber, Kelly de Koning, Roland Kanaar, Joyce HG Lebbink *Journal of Molecular Biology* (2021-07) https://doi.org/gmkjw2 DOI: 10.1016/j.jmb.2021.167058 · PMID: 34023401 · PMCID: PMC8139174

231. **The new SARS-CoV-2 strain shows a stronger binding affinity to ACE2 due to N501Y mutant** Fedaa Ali, Amal Kasry, Muhamed Amin *Medicine in Drug Discovery* (2021-06) https://doi.org/gmk9wv DOI: 10.1016/j.medidd.2021.100086 · PMID: 33681755 · PMCID: PMC7923861

232. **SARS-CoV-2 variants B.1.351 and P.1 escape from neutralizing antibodies** Markus Hoffmann, Prerna Arora, Rüdiger Groß, Alina Seidel, Bojan F Hörnich, Alexander S Hahn, Nadine Krüger, Luise Graichen, Heike Hofmann-Winkler, Amy Kempf, … Stefan Pöhlmann *Cell* (2021-04) https://doi.org/gjnzmz DOI: 10.1016/j.cell.2021.03.036 · PMID: 33794143 · PMCID: PMC7980144

233. **Tracking SARS-CoV-2 variants** https://www.who.int/emergencies/emergency-health-kits/trauma-emergency-surgery-kit-who-tesk-2019/tracking-SARS-CoV-2-variants

234. **SARS-CoV-2 spike P681R mutation, a hallmark of the Delta variant, enhances viral fusogenicity and pathogenicity** Akatsuki Saito, Takashi Irie, Rigel Suzuki, Tadashi Maemura, Hesham Nasser, Keiya Uriu, Yusuke Kosugi, Kotaro Shirakawa, Kenji Sadamasu, Izumi Kimura, … The Genotype to Phenotype Japan (G2P-Japan) Consortium *Cold Spring Harbor Laboratory* (2021-07-19) https://doi.org/gk7d6w DOI: 10.1101/2021.06.17.448820

235. **Diagnostics** COVID-19 Review Consortium *Manubot* (2021-04-30) https://greenelab.github.io/covid19-review/v/32afa309f69f0466a91acec5d0df3151fe4d61b5/#diagnostics

236. **Duration of infectiousness and correlation with RT-PCR cycle threshold values in cases of COVID-19, England, January to May 2020** Anika Singanayagam, Monika Patel, Andre Charlett, Jamie Lopez Bernal, Vanessa Saliba, Joanna Ellis, Shamez Ladhani, Maria

*Virology* (2020-10-26) https://doi.org/gk7fsw DOI: 10.1002/rmv.2184 · PMID: 33105071 · PMCID: PMC7645866

265.     **Turbulent Gas Clouds and Respiratory Pathogen Emissions** Lydia Bourouiba *JAMA* (2020-03-26) https://doi.org/ggqtj4 DOI: 10.1001/jama.2020.4756 · PMID: 32215590

266.     **Clinical characteristics of 24 asymptomatic infections with COVID-19 screened among close contacts in Nanjing, China** Zhiliang Hu, Ci Song, Chuanjun Xu, Guangfu Jin, Yaling Chen, Xin Xu, Hongxia Ma, Wei Chen, Yuan Lin, Yishan Zheng, … Hongbing Shen *Science China Life Sciences* (2020-03-04) https://doi.org/dqbn DOI: 10.1007/s11427-020-1661-4 · PMID: 32146694 · PMCID: PMC7088568

267.     **Evidence for transmission of COVID-19 prior to symptom onset** Lauren C Tindale, Jessica E Stockdale, Michelle Coombe, Emma S Garlock, Wing Yin Venus Lau, Manu Saraswat, Louxin Zhang, Dongxuan Chen, Jacco Wallinga, Caroline Colijn *eLife* (2020-06-22) https://doi.org/gg6dtw DOI: 10.7554/elife.57149 · PMID: 32568070 · PMCID: PMC7386904

268.     **Time Kinetics of Viral Clearance and Resolution of Symptoms in Novel Coronavirus Infection** De Chang, Guoxin Mo, Xin Yuan, Yi Tao, Xiaohua Peng, Fu-Sheng Wang, Lixin Xie, Lokesh Sharma, Charles S Dela Cruz, Enqiang Qin *American Journal of Respiratory and Critical Care Medicine* (2020-05-01) https://doi.org/ggq8xs DOI: 10.1164/rccm.202003-0524le · PMID: 32200654 · PMCID: PMC7193851

269.     **Temporal dynamics in viral shedding and transmissibility of COVID-19** Xi He, Eric HY Lau, Peng Wu, Xilong Deng, Jian Wang, Xinxin Hao, Yiu Chung Lau, Jessica Y Wong, Yujuan Guan, Xinghua Tan, … Gabriel M Leung *Nature Medicine* (2020-04-15) https://doi.org/ggr99q DOI: 10.1038/s41591-020-0869-5 · PMID: 32296168

270.     **COVID-19 and Your Health** CDC *Centers for Disease Control and Prevention* (2020-10-28) https://www.cdc.gov/coronavirus/2019-ncov/prevent-getting-sick/how-covid-spreads.html

271.     **Virological assessment of hospitalized patients with COVID-2019** Roman Wölfel, Victor M Corman, Wolfgang Guggemos, Michael Seilmaier, Sabine Zange, Marcel A Müller, Daniela Niemeyer, Terry C Jones, Patrick Vollmar, Camilla Rothe, … Clemens Wendtner *Nature* (2020-04-01) https://doi.org/ggqrv7 DOI: 10.1038/s41586-020-2196-x · PMID: 32235945

272.     **Clinical predictors and timing of cessation of viral RNA shedding in patients with COVID-19** Cristina Corsini Campioli, Edison Cano Cevallos, Mariam Assi, Robin Patel, Matthew J Binnicker, John C O'Horo *Journal of Clinical Virology* (2020-09) https://doi.org/gg7m96 DOI: 10.1016/j.jcv.2020.104577 · PMID: 32777762 · PMCID: PMC7405830

273.     **Presymptomatic SARS-CoV-2 Infections and Transmission in a Skilled Nursing Facility** Melissa M Arons, Kelly M Hatfield, Sujan C Reddy, Anne Kimball, Allison James, Jesica R Jacobs, Joanne Taylor, Kevin Spicer, Ana C Bardossy, Lisa P Oakley, … John A Jernigan *New England Journal of Medicine* (2020-05-28) https://doi.org/ggszfg DOI: 10.1056/nejmoa2008457 · PMID: 32329971 · PMCID: PMC7200056